\documentclass[10pt,twocolumn,journal]{IEEEtran}
\usepackage{latexsym}
\usepackage{amsfonts}
\usepackage{amsbsy}
\usepackage{amsmath,amssymb}
\usepackage{times}
\usepackage{graphicx}
\usepackage{enumerate}
\usepackage{epstopdf}

\input epsf
\title{ Adaptive Reduced-Rank Minimum Symbol-Error-Rate Receive Processing for Large-Scale Multiple-Antenna Systems}

 \author{Yunlong Cai, Rodrigo C. de Lamare, Benoit Champagne, Boya Qin, and Minjian Zhao
\thanks{

Y. Cai, B. Qin and M. Zhao are with Department of Information Science and Electronic Engineering, Zhejiang University, Hangzhou 310027, China (e-mail: ylcai@zju.edu.cn; qby666@zju.edu.cn; mjzhao@zju.edu.cn).

R. C. de Lamare is with  CETUC-PUC-Rio, 22453-900 Rio de Janeiro, Brazil, and also with the Communications Research Group, Department of Electronics, University of York, YO10 5DD York, U.K. (e-mail: rcdl500@ohm.york.ac.uk).

B. Champagne is with the Department of Electrical and Computer Engineering, McGill University, Montreal, QC, Canada, H3A 0E9 (e-mail: benoit.champagne@mcgill.ca).

This work was supported in part by
the National Natural Science Foundation of China
under Grant  61471319, Zhejiang Provincial Natural Science Foundation
of China under Grant LY14F010013,
the Fundamental Research Funds for the Central Universities, and
 the National High Technology Research and
Development Program (863 Program) of China under Grant 2014AA01A707.
 } }

\begin{document}
\maketitle \thispagestyle{empty}

\vspace{-1cm}
\begin{abstract}
In this work, we propose a novel adaptive reduced-rank receive processing strategy
based on joint preprocessing, decimation and filtering  (JPDF)  for
large-scale  multiple-antenna  systems. In
this scheme, a reduced-rank framework is employed for linear receive
processing and multiuser interference suppression based on the
minimization of the  symbol-error-rate (SER) cost function. We present a
structure with multiple processing branches that performs a
dimensionality reduction, where each branch contains a group of
jointly optimized preprocessing and decimation units, followed by a
linear receive filter. We then develop stochastic gradient (SG)
algorithms to compute the parameters of the preprocessing and
receive filters, along with a low-complexity decimation technique for both binary phase shift keying (BPSK)   and $M$-ary quadrature amplitude modulation (QAM) symbols. In addition, an automatic parameter selection scheme is proposed to further improve the convergence performance of the proposed reduced-rank algorithms.
Simulation results are presented for time-varying wireless environments and
show that the proposed JPDF minimum-SER receive processing strategy and
algorithms achieve a superior performance than existing methods with a reduced computational complexity.
\end{abstract}

\section{Introduction}
\label{sec:intro}

%
The use of large-scale  multiple-antenna  systems, as in e.g., massive
multiple-input multiple-output (MIMO) communications,  has become a highly
popular approach to support the efficient and flexible  multiple access schemes
needed in the next generation of wireless cellular, local area
\cite{marzetta}-\cite{wence} and multibeam satellite networks \cite{arnau}.
%
%
In this context, a key problem
that has been receiving significant attention is the design of efficient and flexible space-time processing techniques at the receiver side.
Space-time processing can be used to separate signals transmitted in the same frequency band, and  provides a practical means of supporting multiple users through space division multiple access  \cite{shengchen}-\cite{sanderson}.
Indeed, systems equipped with large-scale antenna arrays can
substantially increase the system
capacity and improve the quality and reliability of wireless links via beamforming and spatial multiplexing
\cite{largemimo}.
The problem of detecting a desired
user's signal in a  multiuser large-scale  multiple-antenna system presents many signal
processing challenges, including: the need for algorithms with the
ability to process large-dimensional received data vectors, fast and
accurate adjustment of system parameters, scalable computational complexity
and the development of cost-effective interference mitigation
schemes.

Adaptive techniques are among the most commonly used approaches to continually  adjust the receiver weights for detecting a desired signal,  while coping with changes in the radio signal environment
and reducing computational complexity \cite{haykin}.   However, one problem for the standard, i.e. full-rank adaptive algorithms  is that their convergence performance deteriorates rapidly with an increase in  the eigenvalue spread of the received data covariance matrix,  as measured by its condition number \cite{eign}-\cite{MWF}.\footnote{ In the current context of multi-user transceiver design, reduced-rank techniques were primarily meant to reduce the complexity of non-adaptive full-rank algorithms \cite{llyang2008twc} and increase the speed of adaptive parameter estimation tasks. While the computational  complexity of on-line full-rank adaptive algorithms such as  stochastic gradient (SG) based algorithms (e.g., least-mean square (LMS) and its variations) is relatively small, they are characterized by extremely slow convergence in these applications, as will be illustrated in Section \ref{Section6:simulations}. The main problem of interest, therefore, is how to design low-complexity reduced-rank algorithms to increase the  convergence speed and tracking performance in the presence of large-dimensional filters. } This situation is usually worse when a filter with a large number of adaptive weights  is employed to  operate on large-dimensional received data vectors.
In this context, reduced-rank signal processing  has received significant attention in the past years and it has become a  key technique for application to
large adaptive systems, since it can provide faster training,  better
tracking performance and  increased robustness against
interference as compared to standard methods.
The reduced-rank technique projects the large-dimensional received data vector onto a lower dimensional subspace and performs the filtering optimization within this subspace.
A number of reduced-rank algorithms have been developed to design the subspace
projection matrix and the reduced-rank receive filter \cite{eign}-\cite{sabf}.
Among the first schemes are the eigendecomposition-based (EIG)  algorithms
\cite{eign}, \cite{eign2}  and the multistage Wiener filter (MSWF) investigated
in \cite{MWF2}, \cite{MWF}. EIG and MSWF  have fast convergence speed, but
their computational complexity is relatively  high.
 A strategy based on the joint and iterative optimization (JIO) of a subspace projection matrix and a reduced-rank receive filter has been reported in \cite{delamarespl07}-\cite{delamaresp2010}, whereas algorithms using joint preprocessing, decimation and filtering (JPDF) schemes
 have been considered in \cite{delamaretsp2011}-\cite{delamarespl2005}.

However, most of the contributions to date are either based on the minimization
of the mean square error (MSE) and/or the minimum variance criteria
\cite{haykin}-\cite{delamarespl2005}, which are not the most appropriate
metrics from a performance viewpoint in digital communications  where a
standard measure of transmission reliability is the symbol-error-rate (SER)
rather than MSE.  Transceiver design approaches that can minimize the SER have
been reported in  \cite{shengchen}, \cite{mber2}-\cite{euroship} and are termed
adaptive minimum-SER (MSER) techniques. It is also well known that MSER
algorithms lose their advantage when working with large filters and that the
use of reduced-rank techniques can speed up their training. The work in
\cite{mber3} appears to be the first approach to devise a reduced-rank
algorithm with the MSER criterion. However, the resulting scheme is a hybrid
between an EIG or an MSWF approach, and a SER scheme in which only the
reduced-rank receive filter is adjusted in an MSER fashion. The recently
reported work in \cite{caicl2013} investigates a novel adaptive reduced-rank
strategy based on JIO of the receive filters according to the MSER criterion.
The performance results verify that the MSER-JIO reduced-rank algorithm
outperforms the MSER-EIG and the MSER-MSWF reduced-rank algorithms. A
limitation of the work in \cite{caicl2013} is that the subspace projection
matrix might contain a very large number of parameters in large-scale
multiple-antenna systems, which increases the cost and impacts the training of
the receiver. To the best of our knowledge, these existing works on MSER
techniques have not addressed the key problem of performance degradation
experienced when the filters become larger and  their convergence is slow
compared to that of MSE-based techniques.


In this work, we propose a new adaptive reduced-rank receive processing front-end
for multiuser large-scale multiple-antenna systems that incorporates the joint preprocessing,
decimation and filtering (JPDF) structure adaptively optimized on the basis of the MSER criterion.
The proposed
scheme employs a multiple-branch structural framework, where each branch contains a preprocessing filter, a decimation unit and a linear reduced-rank receive filter.
It constitutes a general receive front-end which can be combined with several linear and nonlinear receiver  architecture, but with the key advantage of requiring a much smaller set of adaptive parameters for its adjustment.
 The dimensionality reduction is carried out by the preprocessing filter and the decimation unit. After
dimensionality reduction, a linear receive filter with reduced
dimensionality is applied to suppress the multiuser interference and provide an estimate  of the desired user's symbols.
The final decision is generated among the branch estimates according to the minimum Euclidean distance between a training symbol and the output of each filtering branch.
At each time instant, the preprocessing filter and the reduced-rank receive filter are optimized based on the MSER criterion using the given decimation unit for each branch.
 We devise
stochastic gradient (SG) algorithms to compute the parameters of the
preprocessing and receive filters along with a low-complexity
decimation technique for both binary phase shift keying (BPSK) symbols and $M$-ary quadrature amplitude modulation (QAM) symbols. A unique feature of our scheme is that all
component filters have a small number of parameters and yet can ensure the effectiveness of MSER-type methods.
In order to further improve the convergence performance of the proposed algorithms, we also develop an automatic parameter selection scheme to determine the lengths of the preprocessor and the reduced-rank receive filter.

 The proposed reduced-rank MSER-JPDF technique can operate as a receive-processing front-end which is much simpler than a zero forcing (ZF) or minimum mean square error (MMSE) filter. It can be combined with other interference cancellation algorithms and  offers a significantly better performance than the matched filter.
A detailed analysis of the SG-based adaptive MSER algorithm for updating the parameters of the reduced-rank JPDF structure is carried out   in terms of computational complexity and convergence behavior.
 In simulations over multipath time-varying fading channels,
the proposed MSER-JPDF receive processing strategy and adaptive algorithms
exhibit a performance much superior to that of  full-rank adaptive techniques. Furthermore, compared to  existing MSER-based reduced-rank algorithms,  the new algorithms can significantly reduce the computational complexity and speed up the training. The contributions of this paper are summarized as follows:
\begin{enumerate}[I)~]
\item A novel MSER reduced-rank scheme that incorporates the JPDF structure is proposed for multiuser large-scale multiple-antenna systems.
\item For each branch of the JPDF structure, we develop adaptive SG algorithms to update the preprocessing and receive filters  for BPSK and QAM symbols according to the MSER criterion.
\item  We also propose a selection mechanism for choosing the optimal branch corresponding to the minimum Euclidean distance and present a low-complexity design for the decimation unit.
\item An automatic parameter selection scheme is proposed to further increase the convergence performance of the proposed reduced-rank algorithms.
    \item  For the proposed algorithms, we perform a detailed performance analysis in terms of computational complexity and convergence.
\end{enumerate}

 \newcounter{TempEqCnt}
\setcounter{TempEqCnt}{\value{equation}}
\setcounter{equation}{3}
 \begin{figure*}[!t]
 \begin{equation}
 \mathbf {H}_{\nu, k}(i)= \left( \begin{array}{cccccc} h_{k, \nu, 0}(i) & \ldots & h_{k, \nu, L_p-1}(i) & &  &\\
 & h_{k, \nu, 0}(i-1)&  \ldots  & h_{k, \nu, L_p-1}(i-1) &  & \\
 &  & \ddots  & \ldots & \ddots & \\
& & & h_{k, \nu, 0}(i-P+1) &  \ldots & h_{k, \nu, L_p-1}(i-P+1) \\
\end{array} \right) \label{eq:overallchannelmatrix}
 \end{equation}
 \end{figure*}
 \setcounter{equation}{\value{TempEqCnt}}

The paper is structured as follows. Section \ref{Section2:system}
briefly describes the system model and the problem statement, while
the JPDF reduced-rank scheme is introduced
in Section \ref{Section3:jidf}.
In Section \ref{Section4:proposedalgorithms},
the  proposed adaptive MSER-JPDF reduced-rank algorithms are developed to efficiently update the preprocessing filter and the reduced-rank receive filter for both BPSK and QAM symbols.
The analysis  of  computational complexity and convergence behavior for the proposed algorithms is conducted  in Section \ref{Section5:analysis}.
The simulation results are presented
in Section \ref{Section6:simulations} while conclusions are drawn in Section \ref{Section7:conclusion}.

\section{System Model and Problem Statement}
\label{Section2:system}

We consider the uplink of a large-scale multiple-antenna system with $K$ mobile users and a base station comprised of $L$ identical omnidirectional
antenna elements,
 where $L$ is a large integer and $K\ll L$. The signals from the $K$ users are modulated and transmitted  over multipath channels, after which they are received and demodulated by the base station. In this work, the propagation effects of the multipath channels are modeled by finite impulse response (FIR) filters with $L_p$ coefficients. We assume that the channel can vary  over a block of  transmitted symbols and that the receivers remain perfectly synchronized with the main propagation path.


The demodulated signal received at the $\nu$-th antenna element and $i$-th time instant,
 after applying a filter matched to the signal waveform and sampling at symbol rate, is expressed by
{\small
\begin{equation}
r_{\nu}(i)=\sum^{K-1}_{k=0}\sum^{L_p-1}_{\mu=0}h_{k,\nu,\mu}(i)b_k(i-\mu)+n_{\nu}(i), \quad \nu\in\{0, \ldots, L-1\},
\end{equation}
}
where $h_{k,\nu,\mu}(i)$ is the sampled impulse response between user $k$ and receive antenna $\nu$ for path $\mu\in\{0,\ldots, L_p-1\}$,  $b_k(i)$ are  the data symbols of user $k\in \{0, \ldots, K-1\}$, and $n_{\nu}(i)$ are samples of white Gaussian noise. By collecting the samples of the received signal and organizing them in a window of $P$ symbols
 for each antenna element, we obtain the $LP\times 1$ received vector as
\begin{equation}
\mathbf{r}(i)=\mathbf{H}(i)\mathbf{b}(i)+\mathbf{n}(i).\label{eq:systemmodel12}
\end{equation}
 In this expression, $\mathbf{r}(i)=[\mathbf{r}^{T}_0(i),\ldots, \mathbf{r}^{T}_{L-1}(i)]^{T}$ contains the signals that are collected by the $L$ antennas, the $P\times 1$ vector $\mathbf{r}_{\nu}(i)=[r_{\nu}(i), \ldots, r_{\nu}(i-P+1)]^{T}$, contains the signals that are collected by the $\nu$-th antenna and are organized into a vector.
 The $K(P+L_p-1)\times 1$ vector $\mathbf{b}(i)=[\mathbf{b}^{T}_0(i), \ldots, \mathbf{b}^{T}_{K-1}(i)]^T$ is composed of the data symbols that are transmitted from the $K$ users, with $\mathbf{b}_k(i)=[b_k(i),\ldots, b_k(i-(P+L_p-2))]^T$ being the $i$-th block of transmitted symbols with dimensions $(P+L_p-1)\times 1$. The $LP\times K(P+L_p-1)$ channel matrix $\mathbf{H}(i)$ is expressed as
  {\small
  \begin{equation}
 \mathbf {H}(i)= \left( \begin{array}{cccc} \mathbf{H}_{0, 0}(i) & \mathbf{H}_{0, 1}(i) & \ldots & \mathbf{H}_{0, K-1}(i) \\
\mathbf{H}_{1, 0}(i) & \mathbf{H}_{1, 1}(i)  & \ldots & \mathbf{H}_{1, K-1}(i)\\
\vdots & \vdots & \ddots & \vdots \\
\mathbf{H}_{L-1, 0}(i) & \mathbf{H}_{L-1, 1}(i) & \ldots & \mathbf{H}_{L-1, K-1}(i) \\
\end{array} \right)
 \end{equation}
 }
 where the $P\times (P+L_p-1)$ matrices $\mathbf {H}_{\nu, k}(i)$ are given by (\ref{eq:overallchannelmatrix}).
Specifically, let us assume that the sequence of transmitted symbols
by the $k$-th users, i.e., $b_k(i)$,
are independent and identically
distributed (i.i.d) random variables drawn from a given symbol set.
In this work,  BPSK  and  $M$-ary square QAM symbol constellations are adopted, although extensions to other types of constellations are possible. For the BPSK case,  the symbols $b_k(i)$ are  uniformly drawn from $\{\pm 1\}$. For the $M$-ary square QAM case, the symbols are uniformly drawn from $\{F_m+jF_n: 1\leq m, n \leq \sqrt{M}\}$, where integer $M$  is a perfect square and  we define $F_n=2n-\sqrt{M}-1$.
Without loss of generality, we index the desired user as $k=0$.
The term $\mathbf{n}(i) \in \mathbb{C}^{LP\times 1}$ is an  additive noise vector,
 which is assumed to be Gaussian, spatially white, independent over time, with zero-mean and covariance matrix $E[\mathbf{n}(i)\mathbf{n}^{H}(i)] = \sigma^2 \mathbf{I}$,  where $\sigma^2$   denotes
the variance, $\mathbf{I}$ is an identity matrix of dimension $LP$, and $(.)^{H}$ stands for the Hermitian transpose operation.
The transmitted symbols  from the different users and the noise vectors are mutually independent.

The full-rank  beamforming receiver design is equivalent to determining an FIR filter
$\mathbf{w}(i)$
with $LP$ coefficients that provide an estimate of the desired signal as
\setcounter{TempEqCnt}{\value{equation}}
\setcounter{equation}{4}
\begin{equation}
\hat{b}_0(i)=\mathcal{Q}\{\mathbf{w}^{H}(i)\mathbf{r}(i)\},
\end{equation}
where $\mathcal{Q}\{.\}$ represents the quantization operation for the given constellation and
$\mathbf{w}(i)=[w_{0}, \ldots, w_{LP-1}]^{T}\in \mathbb{C}^{LP \times 1}$ is the complex weight vector of the receive filter.
However, the  dimensionality of $\mathbf{w}(i)$ can become excessive for large antenna-array systems, which leads to computationally intensive implementations and slow convergence
performance when full-rank adaptive algorithms are employed \cite{eign}-\cite{MWF}.
The  reduced-rank schemes, which process the received vector $\mathbf{r}(i)$ in two stages, have been proposed to overcome these limitations \cite{eign}-\cite{delamaresp2010}.
The first stage performs  a dimensionality reduction by projecting the large dimensional data vector $\mathbf{r}(i)$ onto a lower dimensional subspace.
The second stage is carried out by a reduced-rank receive filter. The output of a reduced-rank scheme is given by
\begin{equation}
\hat{b}_0(i)=\mathcal{Q}\{\mathbf{\bar{w}}^{H}(i)\mathbf{S}^{H}_{D}(i)\mathbf{r}(i)\}=\mathcal{Q}\{\mathbf{\bar{w}}^{H}(i)\mathbf{\bar{r}}(i)\}, \label{eq:detectsymbol2014}
\end{equation}
where  $\mathbf{S}_{D}(i) $ denotes an $LP \times D$ projection matrix\footnote{In this paper, projection simply refers to a linear transformation from a space of large dimension $LP$ to a space of smaller dimension $D$.} which is applied to the received vector to extract the most
important information from the data by performing dimensionality reduction, where $1 \leq D < LP$, and
$\mathbf{\bar{w}}(i)=[\bar{w}_0,\bar{w}_{1},\ldots,\bar{w}_{D-1}]^{T} \in \mathbb{C}^{D \times 1}$ denotes the  reduced-rank receive filter.
In (\ref{eq:detectsymbol2014}), for convenience, the $D \times 1$ projected received vector is denoted as
 \begin{equation}
\mathbf {\bar{r}}(i)=\mathbf{S}^{H}_{D}(i)\mathbf {r}(i).
\end{equation}

The basic problem in implementing the MSER reduced-rank receive processing scheme
 is how to effectively devise the projection matrix $\mathbf{S}_{D}(i)$ and the reduced-rank receive filter $\mathbf{\bar{w}}(i)$. In the following sections, we propose and investigate a new  structure and adaptive algorithms for on-line estimation of these quantities based on the minimization of the SER cost function \cite{shengchen}, \cite{mber2}-\cite{euroship}.

\section{Proposed MSER-JPDF Reduced-Rank Linear Receive Processing Scheme}
\label{Section3:jidf}

In this section, we detail the proposed MSER-JPDF reduced-rank linear
receive processing scheme.
Apart from the conventional reduced-rank techniques, the most direct method to reduce the dimensionality of the received vector is
to decimate its content, i.e., to retain a subset of its elements while discarding the rest.
However,
this approach may entail a loss of information and therefore result in poor  convergence performance.
 To overcome this problem, the proposed  technique first performs a linear preprocessing operation on the received vector,
then the output of the preprocessor is handled by a decimator, followed by a reduced-rank receive filter \cite{delamaretvt10}, \cite{delamarespl2005}.
With the aid of the linear preprocessing, essential information conveyed by the input signals can  be preserved  in the lower dimensional data vector operated upon by the reduced-rank receive filter.
Since wireless channels  tend to vary rapidly, and the determination of the optimal decimator is prohibitively complex  due to the need for an exhaustive search, we propose to create several branches of preprocessing, decimation and reduced-rank receive filters based on a number of different fixed decimators \cite{delamaretsp2011}.
For a given branch, the preprocessor and the reduced-rank receive filter are jointly designed according to the MSER criterion.
The final symbol estimate is generated  among the outputs of the multiple filtering branches according to the minimum Euclidean distance criterion.


%
%

\subsection{Overview of the MSER-JPDF Scheme}

\begin{figure}[!hhh]
\centering \scalebox{0.45}{\includegraphics{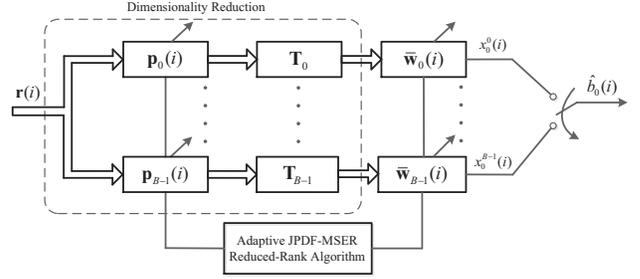}}
\caption{Structure of the proposed MSER-JPDF scheme}
\label{fig:jidfmber}
\end{figure}

We design the subspace projection matrix $\mathbf{S}_{D}(i)$ by
considering preprocessing and decimation. In this case, the receive
filter length is substantially decreased (i.e., $D \ll L$), which  in turn
significantly reduces the computational complexity and leads to very fast
training for large-scale multiple-antenna systems. The proposed
MSER-JPDF scheme for  the desired user is depicted in Fig.
\ref{fig:jidfmber}. The $LP \times1$ received vector $\mathbf{r}(i)$
is processed by a parallel structure consisting of $B$ branches, where each branch
contains a preprocessing filter and a decimation unit, followed by a reduced-rank receive filter. In the $l$-th branch, $l \in \{0, \ldots, B-1\}$, the received vector is
operated by the preprocessor
$\mathbf{p}_{l}(i)=[p_{0,l}(i),\ldots, p_{I-1,l}(i)]^{T}$ with
 length  $I < LP$. The output of the preprocessor on the
$l$-th branch is expressed as
\begin{equation}
\mathbf{\tilde{r}}_{l}(i)=\mathbf{P}^{H}_{l}(i)\mathbf{r}(i)
\end{equation}
where the $LP\times LP$ Toeplitz convolution matrix $\mathbf{P}_{l}(i)$ is given by\footnote{ Space-time processing is used in the current setting due to both intersymbol and multiuser interferences. In this case, these interferences can be jointly suppressed by jointly processing the received signal in the temporal and spatial domains \cite{delamaretvt2011ST}, \cite{IETJIDF}.}
\begin{equation}
\mathbf {P}_{l}(i)= \left( \begin{array}{cccc} p_{0,l}(i) & 0 & \ldots & 0 \\
\vdots & p_{0,l}(i)  & \ldots & 0\\
p_{I-1,l}(i) & \vdots & \ldots &0 \\
0 & p_{I-1,l}(i) & \ldots & 0 \\
\vdots & \vdots & \ddots & \vdots\\
0 & 0 & \ldots &  p_{0,l}(i)
\end{array} \right).
\end{equation}
In order to facilitate the description of the scheme, we describe the vector $\tilde{\bf r}_l(i)$ as a function of the preprocessor vector $\mathbf{p}_l(i)$ and an input data matrix $\mathbf{R}(i)$ as follows:
\begin{equation}
\mathbf{\tilde{r}}_{l}(i)=\mathbf{P}^{H}_{l}(i)\mathbf{r}(i)=\mathbf{R}(i)\mathbf{p}^{*}_{l}(i)
\end{equation}
where $(.)^{*}$ stands for the element-wise conjugate operation and
the $LP\times I$ matrix $\mathbf{R}(i)$ is obtained from the received samples
$\mathbf{r}(i)=[\mathbf{r}^{T}_0(i),\ldots, \mathbf{r}^{T}_{L-1}(i)]^{T}$ and has a Hankel
structure shown in \cite{delamaretsp2011}, \cite{delamarespl2005} and \cite{hankel}.

The dimensionality reduction is performed by a decimation unit implemented as a
$D\times LP$ decimation matrix $\mathbf{T}_{l}$ that projects
$\mathbf{\tilde{r}}_{l}(i)$ onto the $D\times 1$ vector
$\mathbf{\bar{r}}_{l}(i)$, where $D$ is the
rank of $\mathbf{T}_l$. Specifically, the $D \times 1$ vector $\mathbf{\bar{r}}_{l}(i)$ for the
$l$-th branch is given by
\begin{equation}
\mathbf{\bar{r}}_{l}(i)=\underbrace{\mathbf{T}_{l}\mathbf{P}^{H}_{l}(i)}_{\mathbf{S}^{H}_{D,l}(i)}\mathbf{r}(i)=\mathbf{T}_{l}\mathbf{\tilde{r}}_{l}(i)=\mathbf{T}_{l}\mathbf{R}(i)\mathbf{p}^{*}_{l}(i)\label{eq:lineartransform}
\end{equation}
where $\mathbf{S}_{D,l}(i)$ denotes the equivalent subspace
projection matrix.  The output of
the reduced-rank receive filter $\mathbf{\bar{w}}_l(i)$ corresponding to the $l$-th branch
is given by
\begin{equation}
{x}^{l}_{0}(i)=\mathbf{\bar{w}}^{H}_l(i)\mathbf{\bar{r}}_{l}(i)\label{eq:outputeachbranch}
\end{equation}
which is used in the minimization of the error probability for that
branch.

As seen from Fig. \ref{fig:jidfmber}, the proposed scheme employs $B$ parallel branches of preprocessors, decimators and receive filters.
The optimum branch  is selected according to the minimum Euclidean
distance criterion, that is:
\begin{equation}
l_{opt}=\arg \min_{0\leq l \leq B-1} |b_0(i)-{x}^{l}_{0}(i)|,\label{eq:errorpl}
\end{equation}
where $|.|$ represents the magnitude of a complex scalar and $b_0(i)$ refers to a known  sequence of training symbols transmitted by  the desired user and available at the receiver side.
The  output of the MSER-JPDF is given by
\begin{equation}
\hat{b}_{0}(i)=\mathcal{Q}\{{x}^{l_{opt}}_{0}(i)\}=\mathcal{Q}\{\mathbf{\bar{w}}^{H}_{l_{opt}}(i)\mathbf{\bar{r}}_{l_{opt}}(i)\}.
\end{equation}

We can claim that more branches will result in better performance for the proposed algorithm. However, considering the affordable complexity, we have to configure the algorithm with the number of branches as small as possible and yet achieve a satisfactory performance. As will be shown in the simulation results, the proposed MSER-JPDF reduced-rank algorithm with the number of branches $B$ set to $4$ offers an attractive tradeoff between  performance and  complexity.

\subsection{Design of the Decimation Unit}

In this work, the elements of the decimation matrix only take the value $0$ or $1$, which corresponds to the decimation unit simply keeping or discarding its samples.
The optimal decimation scheme exhaustively explores all possible patterns which select $D$ samples out of $LP$ samples.
In this case, the design can be viewed as a combinatorial problem where the total number of patterns is $B=LP (LP-1)\ldots(LP-D+1)$.
 However, the optimal decimation scheme is too complex for practical use and several low-complexity  approaches such as the uniform decimation, the pre-stored decimation and the random decimation  have been investigated \cite{rodrigotsp2009JIO}, \cite{delamaretsp2011}. Among these approaches, the pre-stored decimation provides a suboptimal solution
  to generate  simple and yet effective decimation matrices.
In this work, we therefore use the low-complexity  pre-stored method, in which the $l$-th decimation matrix is formed as
\begin{equation}
 \mathbf{T}_{l}=\left[ \begin{array}{cccc} \mathbf{t}_{l,0} &  \mathbf{t}_{l,1}  & \ldots  & \mathbf{t}_{l,D-1}
 \end{array} \right]^{T}, \quad l\in\{0,\ldots, B-1\}
\end{equation}
where the $LP\times 1$ vector $\mathbf{t}_{l,d}$   is composed of a single $1$ and $LP-1$ $0$s, as given by
\begin{equation}
\mathbf{t}_{l,d}= [\underbrace{0, \ldots, 0}_{v_{l,d}}, 1, \underbrace{0, \ldots, 0}_{LP-v_{l,d}-1}]^{T}, \quad d \in\{0,\ldots, D-1\}
\end{equation}
and $v_{l,d}$ is the number of zeros before the non-zero entry.
 Hence, each row of decimation matrix $\mathbf{T}_l$ is all zero but for one entry which is set to $1$; furthermore, we require that the ones in different row occupy different positions.
Specifically, we set the values of $v_{l,d}$ in a deterministic way, which can be expressed as
\begin{equation}
v_{l,d}=\bigg\lfloor{\frac{LP}{D}\bigg\rfloor} d+l
\end{equation}
where $\lfloor{.\rfloor}$ represents the floor function, which returns the largest integer that is smaller than or equal to its argument.
%
The simulation results will show that the proposed MSER-JPDF reduced-rank
scheme with the above suboptimal decimation scheme performs very well.

\setcounter{TempEqCnt}{\value{equation}}
\setcounter{equation}{23}
\begin{figure*}[!t]
\begin{equation}
\mathcal{P}^{l}_{e}(\mathbf{\bar{w}}_l(i),\mathbf{p}_{l}(i))=\left
 \{
\begin{array}{cc}
\int^{0}_{-\infty}\frac{1}{\sqrt{2\pi}\rho}e^{-\frac{|x-\Re[{x}^{l}_0(i)]|^2}{2\rho^2}}dx,  \quad \textrm {if} \quad b_0(i)=+1 \\
\int^{\infty}_{0}\frac{1}{\sqrt{2\pi}\rho}e^{-\frac{|x-\Re[{x}^{l}_0(i)]|^2}{2\rho^2}}dx,   \quad  \textrm {if} \quad  b_0(i)=-1 \\
 \end{array}
\right.
\label{eq:SERfunctionbpsk1}
\end{equation}
 \end{figure*}
\setcounter{equation}{\value{TempEqCnt}}

\section{Proposed Adaptive  Algorithms}
\label{Section4:proposedalgorithms}



In this section, we introduce the proposed adaptive MSER-JPDF reduced-rank algorithms. In particular,
we develop the MSER-based adaptive  algorithms to
update the preprocessing filter $\mathbf{p}_l(i)$ and the reduced-rank receive filter $\bar{\mathbf{w}}_l(i)$ for each
branch.  Note that the subspace projection introduced in (\ref{eq:lineartransform}) is linear  and hence,  the transformed noise,  i.e. $\mathbf{T}_l \mathbf{P}_l^H(i) \mathbf{n}(i)$, also follows a Gaussian distribution, allowing us to derive the MSER based algorithms. We first consider the case of BPSK symbols, and then extend the presentation to QAM symbols. The derivation is general in its conceptual approach and it could be extended to other modulation schemes. Subsequently, we propose an automatic parameter selection scheme to determine the lengths of the preprocessor and the reduced-rank receive filter during on-line operation.

\subsection{Adaptive MSER-JPDF Algorithm for BPSK Symbols}

\begin{table*}[t]
\centering
 \caption{\normalsize  Adaptive MSER-JPDF reduced-rank algorithm for BPSK Symbols} {
\begin{tabular}{ll}
\hline
 $1$ & Set step-size values $\mu_{w}$ and $\mu_{S_{D}}$ and the number of branches $B$.\\
 $2$ & Initialize $\mathbf {\bar{w}}_l(0)$ and $\mathbf {p}_{l}(0)$. Set $\mathbf{T}_{0},\ldots, \mathbf{T}_{B-1}$.\\
 $3$ &   for  time instant $i\in \{0, 1, \ldots \}$ do \\
 $4$ & ~ for $l \in \{0,\ldots,B-1\}$ do\\
 $5$ & ~~~  Compute  $\mathbf {p}_{l}(i+1)$ based on (\ref{eq:adaptiveSGbpsk2}) using $\mathbf{\bar{w}}_l(i)$ and $\mathbf{T}_{l}$. \\
  $6$ & ~~~  Compute $\mathbf{\bar{w}}_l(i+1)$ based on (\ref{eq:adaptiveSGbpsk1}) using $\mathbf {p}_{l}(i+1)$ and $\mathbf{T}_{l}$.  \\
 $7$ & ~ end  \\
$8$ & ~ Select the optimal branch $l_{opt}=\arg \min_{0\leq l \leq B-1} |b_0(i)-{x}^{l}_{0}(i)|$, where ${x}^{l}_{0}(i)=\mathbf{\bar{w}}^{H}_l(i+1)\mathbf{T}_{l}\mathbf{P}^{H}_{l}(i+1)\mathbf{r}(i)$. \\
     & ~ Generate the estimate symbol corresponding to branch $l_{opt}$: $\hat{b}_0(i)=\mathcal{Q}\{{x}^{l_{opt}}_{0}(i)\}$.\\
\hline
\label{tab:table1}
\end{tabular}
}
\end{table*}

Firstly, let us consider the case with BPSK symbols.
The symbol decision $\hat{b}_0(i)$ can be made as
\begin{equation}
\hat{b}_0(i)=\left
 \{
\begin{array}{cc}
+1,  \quad \textrm {if} \quad \Re[{x}^{l}_0(i)] \geq 0 \\
 -1,   \quad  \textrm {if} \quad \Re[{x}^{l}_0(i)] < 0 \\
 \end{array}
\right.
\label{eq:QPSKdecisionbpsk}
\end{equation}
where $\Re[.]$ selects the real part and $x_0^l(i)$ is the output of the $l$-th branch as given in (\ref{eq:outputeachbranch}).
 Note that at the $i$-th block of transmitted symbols and for a given desired symbol $b_{0}(i)$, there are $N_b=2^{(P+L_p-1)K-1}$ different possible arrangements (i.e. $((P+L_p-1)K-1)$-tuples) of the binary multiuser interference (MUI) symbols $b_{k}(i-\mu)$, where $k\in\{1, \ldots, K-1\}$ and $\mu \in \{0, \ldots, P+L_p-2\}$, and the binary intersymbol interference (ISI) symbols $b_{0}(i-\mu^{'})$, where $\mu^{'} \in \{1, \ldots, P+L_p-2\}$.
Then, we define a set:
\begin{equation}
\mathcal{X}=\{\mathbf{\tilde{b}}^{0}, \mathbf{\tilde{b}}^{1}, \ldots, \mathbf{\tilde{b}}^{N_b-1}\}
\end{equation}
which contains all  possible  transmitted symbol vectors of size $K(P+L_p -1)\times 1$ for a given $b_0(i)$.
The noise-free component of the reduced-rank receive filter corresponding to the $l$-th branch takes its value from the set 
\begin{equation}
\mathcal{Y}=\{\bar{x}^{l,q}_0=\mathbf{\bar{w}}^{H}_l(i)\mathbf{T}_{l}\mathbf{P}^{H}_l(i)\mathbf{H}(i)\mathbf{\tilde{b}}^{q}:\ 0\leq q \leq N_b-1\}
\end{equation}
where $\mathbf{\tilde{b}}^{q}\in \mathcal{X}$.
Based on the  Gaussian distribution for the additive noise in (\ref{eq:systemmodel12}) and assuming an equiprobable model for the $N_b$ possible transmit vectors $\tilde{\mathbf{b}}^q$, we  can express the  probability density function (PDF) for the real part of the reduced-rank receive filter output corresponding to the $l$-th branch,  conditioned on the desired symbol $b_{0}(i)$, as
\begin{equation}
\begin{split}
f_l(x|b_{0}(i))&=\frac{1}{ N_b \sigma\sqrt{2\pi \mathbf{\bar{w}}^{H}_l(i)\mathbf{T}_l\mathbf{P}^{H}_l(i)\mathbf{P}_l(i)\mathbf{T}^{H}_l\mathbf{\bar{w}}_l(i)}}\\&\quad\times\sum^{N_b-1}_{q=0}e^{-\frac{|x-\Re[\bar{x}^{l,q}_0]|^2}{2\sigma^2\mathbf{\bar{w}}^{H}_l(i)\mathbf{T}_l\mathbf{P}^{H}_l(i)\mathbf{P}_l(i)\mathbf{T}^{H}_l\mathbf{\bar{w}}_l(i)} }\label{eq:conditionalpdf}
\end{split}
\end{equation}
where $\bar{x}^{l,q}_0\in \mathcal{Y}$.
In practice, the PDF of the reduced-rank receive filter output  should be estimated using kernel density estimation \cite{densityestimation1} based on a block of experimental data. Specifically, given $J$ observations of the array output vector, denoted as $\mathbf{r}^{\nu}$ where $\nu \in\{0,1,\ldots, J-1\}$, a kernel estimate of the true PDF is given by
\begin{equation}
\begin{split}
\tilde{f}_l(x|b_{0}(i))&=\frac{1}{J\rho\sqrt{2\pi \mathbf{\bar{w}}^{H}_l(i)\mathbf{T}_l\mathbf{P}^{H}_l(i)\mathbf{P}_l(i)\mathbf{T}^{H}_l\mathbf{\bar{w}}_l(i)}}\\&\quad\times\sum^{J-1}_{\nu=0}e^{-\frac{|x-\Re[{x}^{l,\nu}_0]|^2}{2\rho^2\mathbf{\bar{w}}^{H}_l(i)\mathbf{T}_l\mathbf{P}^{H}_l(i)\mathbf{P}_l(i)\mathbf{T}^{H}_l\mathbf{\bar{w}}_l(i)} }\label{eq:conditionalpdftraining}
\end{split}
\end{equation}
where $\rho$ represents the radius parameter or width of the kernel density estimate and $x^{l,\nu}_0$ denotes the output of the $l$-th branch reduced-rank receive filter corresponding to the  $\nu$-th testing vector $\mathbf{r}^{\nu}$.
In order to design a practical on-line adaptive algorithm with moderate complexity, we consider the following single-sample estimate of the true PDF\footnote{As discussed in \cite{shengchen}, \cite{shengchentsp2004}, \cite{shengchensp2008}, the SER does not change significantly with the quantity $\mathbf{\bar{w}}^{H}_l(i)\mathbf{T}_l\mathbf{P}^{H}_l(i)\mathbf{P}_l(i)\mathbf{T}^{H}_l\mathbf{\bar{w}}_l(i)$. Therefore, in order to simplify the gradient of the resulting estimated SER,
we can drop the term $\mathbf{\bar{w}}^{H}_l(i)\mathbf{T}_l\mathbf{P}^{H}_l(i)\mathbf{P}_l(i)\mathbf{T}^{H}_l\mathbf{\bar{w}}_l(i)$ and employ a constant kernel width $\rho$, which leads to significant reductions in computational complexity.
}
\begin{equation}
\tilde{f}_l(x|b_{0}(i))=\frac{1}{\sqrt{2 \pi}\rho}e^{-\frac{|x-\Re[{x}^{l}_0(i)]|^2}{2\rho^2}}.\label{eq:singlesampleestimatepdf}
\end{equation}
Subsequently, based on the  PDF estimate (\ref{eq:singlesampleestimatepdf}) we obtain the instantaneous
SER estimate
corresponding to the $l$-th branch of the proposed reduced-rank scheme, as given by (\ref{eq:SERfunctionbpsk1}).
In order to simplify this expression, we define a new quantity $s=\frac{x-\Re[{x}^{l}_0(i)]}{\sqrt{2}\rho}$. Then,
%
we can express the BPSK   SER by means of the following unified expression:
\setcounter{TempEqCnt}{\value{equation}}
\setcounter{equation}{24}
\begin{equation}
\mathcal{P}^{l}_{e}(\mathbf{\bar{w}}_l(i),\mathbf{p}_{l}(i))=\int^{\frac{-\Re[{x}^{l}_0(i)]\textup{sign}(b_0(i))}{\sqrt{2}\rho}}_{-\infty}\frac{1}{\sqrt{\pi}}e^{-s^{2}}ds\label{eq:SERfunctionbpsk3}
\end{equation}
where $\textup{sign}(.)$ denotes the signum function.

In the following, we  derive the gradient terms for the reduced-rank receive filter and the preprocessing vector for each branch.
By computing the gradient of (\ref{eq:SERfunctionbpsk3})
with respect to $\mathbf {\bar{w}}^{*}_l(i)$ and after further
mathematical manipulations we obtain
\begin{equation}
\begin{split}
\frac{\partial \mathcal{P}^{l}_{e}(\mathbf{\bar{w}}_l(i),\mathbf{p}_{l}(i))}{\partial \mathbf{\bar{w}}^{*}_l} &
=-\textup{sign}(b_0(i))\frac{1}{\sqrt{2\pi}\rho}e^{- \frac{|\Re[{x}^{l}_0(i)]|^2}{2\rho^2}}\\&\quad\times\mathbf{T}_l\mathbf{P}^{H}_l(i)\mathbf{r}(i)
\label{eq:instantSGbpsk1}.
\end{split}
\end{equation}
In order to derive the gradient terms for the preprocessor
$\mathbf{p}_{l}(i)$, we define the $LP\times 1$ vector
$\mathbf{T}^{H}_l\mathbf{\bar{w}}_l(i)=[d_0(i), d_1(i),\ldots, d_{LP-1}(i)]^{T}$.
The reduced-rank receive filter output can be rewritten  as
\begin{equation}
{x}^{l}_0(i)=\mathbf{\bar{w}}^{H}_l(i)\mathbf{T}_{l}\mathbf{P}^{H}_l(i)\mathbf{r}(i)=\mathbf{p}^{H}_{l}(i)\mathbf{D}^{H}(i)\mathbf{r}(i),
\end{equation}
where the $LP\times I$ matrix $\mathbf{D}(i)$ has the following structure
\begin{equation}
\mathbf{D}(i)= \left( \begin{array}{cccc} d_{0}(i) & 0 & \ldots & 0 \\
d_1(i) & d_0(i) & \ldots & 0\\
\vdots & \vdots & \ddots & \vdots \\
d_{I-1}(i) & d_{I-2}(i) & \ldots & d_0(i)\\
d_{I}(i) & d_{I-1}(i) & \ldots & d_1(i)\\
\vdots & \vdots & \ddots & \vdots\\
d_{LP-1}(i) & d_{LP-2}(i) & \ldots & d_{LP-I}(i)
\end{array} \right).
\end{equation}
 By computing the gradient of (\ref{eq:SERfunctionbpsk3}) with respect to  $\mathbf{p}^{*}_l(i)$, we obtain
 \begin{equation}
 \begin{split}
 \frac{\partial \mathcal{P}^{l}_{e}(\mathbf{\bar{w}}_l(i),\mathbf{p}_{l}(i))}{\partial \mathbf{p}^{*}_l} &
=-\textup{sign}(b_0(i))\frac{1}{\sqrt{2\pi}\rho}e^{- \frac{|\Re[{x}^{l}_0(i)]|^2}{2\rho^2}}\\&\quad\times\mathbf{D}^{H}(i)\mathbf{r}(i)
\label{eq:instantSGbpsk2}.
\end{split}
 \end{equation}

The preprocessing filter and the reduced-rank
receive filter  are jointly optimized
according to the MSER criterion \cite{shengchen}, \cite{shengchensp2008}.  The proposed  SG update equations for the $l$-th branch for BPSK symbols are
obtained by substituting the gradient terms (\ref{eq:instantSGbpsk1}) and (\ref{eq:instantSGbpsk2}) in the following expressions
\begin{equation}
\mathbf{\bar{w}}_l(i+1)=\mathbf{ \bar{w}}_l(i)-\mu_{w}\frac{\partial \mathcal{P}^{l}_{e}(\mathbf{\bar{w}}_l(i),\mathbf{p}_{l}(i))}{\partial \mathbf{\bar{w}}^{*}_l}\label{eq:adaptiveSGbpsk1}
\end{equation}
and
\begin{equation}
\mathbf{p}_{l}(i+1)=\mathbf{p}_{l}(i)-\mu_{p}\frac{\partial \mathcal{P}^{l}_{e}(\mathbf{\bar{w}}_l(i),\mathbf{p}_{l}(i))}{\partial \mathbf{p}^{*}_l},\label{eq:adaptiveSGbpsk2}
\end{equation}
where $\mu_{w}$ and $\mu_{p}$ are the step-size values. At each time instant, these two vectors for a given branch $l$ are updated in an alternating way.
The algorithm is devised to
start its operation in the training (TR) mode, where a known training sequence $b_0(i)$ is employed,
and then to switch to
the decision-directed (DD) mode, where the estimated symbols $\hat{b}_0(i)$ from (\ref{eq:QPSKdecisionbpsk}) are used to compute  the gradients in (\ref{eq:instantSGbpsk1}) and (\ref{eq:instantSGbpsk2}).
 Expressions
(\ref{eq:adaptiveSGbpsk1}) and (\ref{eq:adaptiveSGbpsk2}) need initial values, that is
$\mathbf {\bar{w}}_l(0)$ and $\mathbf{p}_{l}(0)$.
The
proposed MSER-JPDF algorithm for  BPSK modulation is summarized in Table
\ref{tab:table1}.

\subsection{Adaptive MSER-JPDF Algorithm for QAM Symbols}

\setcounter{TempEqCnt}{\value{equation}}
\setcounter{equation}{36}
\begin{figure*}[!t]
\begin{equation}
\hat{b}^{R}_0(i)=\left
 \{
\begin{array}{cl}
F_1, & \textrm {if}~~\Re[{x}^{l}_0(i)] \leq \omega^{l}_{0, 0}(i)(F_1+1) \\
 F_m,  &   \textrm {if}~~\omega^{l}_{0, 0}(i)(F_m-1)<\Re[{x}^{l}_0(i)] \leq \omega^{l}_{0, 0}(i)(F_m+1),~2\leq m \leq \sqrt{M}-1\\
 F_{\sqrt{M}}, &  \textrm{if}~~\Re[{x}^{l}_0(i)] > \omega^{l}_{0, 0}(i)(F_{\sqrt{M}}-1)
 \end{array}
\right.
\label{eq:QAMdecisionreal}
\end{equation}
\end{figure*}
\setcounter{equation}{\value{TempEqCnt}}

\setcounter{TempEqCnt}{\value{equation}}
\setcounter{equation}{37}
\begin{figure*}[!t]
\begin{equation}
\hat{b}^{I}_0(i)=\left
 \{
\begin{array}{cl}
F_1, & \textrm {if}~~\Im[{x}^{l}_0(i)] \leq \omega^{l}_{0, 0}(i)(F_1+1) \\
 F_n,  &  \textrm {if}~~\omega^{l}_{0, 0}(i)(F_n-1)<\Im[{x}^{l}_0(i)] \leq \omega^{l}_{0, 0}(i)(F_n+1),~2\leq n \leq \sqrt{M}-1\\
 F_{\sqrt{M}}, & \textrm{if}~~\Im[{x}^{l}_0(i)] > \omega^{l}_{0, 0}(i)(F_{\sqrt{M}}-1)
 \end{array}
\right.
\label{eq:QPSKdecisionimag}
\end{equation}
\end{figure*}
\setcounter{equation}{\value{TempEqCnt}}

Let us consider the case with $M$-ary  QAM symbols.
For the proposed reduced-rank scheme, the symbol error probability regarding the $l$-th branch can be represented as
\begin{equation}
\begin{split}
\mathcal{P}^{l}_{e}(\mathbf{\bar{w}}_l(i),\mathbf{p}_{l}(i))&=\mathcal{P}^{l,R}_{e}(\mathbf{\bar{w}}_l(i),\mathbf{p}_{l}(i))+\mathcal{P}^{l,I}_{e}(\mathbf{\bar{w}}_l(i),\mathbf{p}_{l}(i))\\&\quad-\mathcal{P}^{l,R}_{e}(\mathbf{\bar{w}}_l(i),\mathbf{p}_{l}(i))\mathcal{P}^{l,I}_{e}(\mathbf{\bar{w}}_l(i),\mathbf{p}_{l}(i))
\end{split}
\end{equation}
where $\mathcal{P}^{l}_{e}(\mathbf{\bar{w}}_l(i),\mathbf{p}_{l}(i))=\textup{Prob}\{b_{0}(i)\neq\hat{b}_0(i)\}$ denotes the total SER while $\mathcal{P}^{l,R}_{e}(\mathbf{\bar{w}}_l(i),\mathbf{p}_{l}(i))=\textup{Prob}\{\Re[b_{0}(i)]\neq \Re[\hat{b}_0(i)]\}$ and $\mathcal{P}^{l,I}_{e}(\mathbf{\bar{w}}_l(i),\mathbf{p}_{l}(i))=\textup{Prob}\{\Im[b_{0}(i)]\neq \Im[\hat{b}_0(i)]\}$ denote the real part and imaginary part SERs, respectively (here,  $\Im[.]$ selects  the imaginary part). The optimization problem is formulated to minimize an upper bound of the SER as follows \cite{shengchen}:
\begin{equation}
\min_{\mathbf{\bar{w}}_l(i), \mathbf{p}_{l}(i)} \mathcal{\bar{P}}^{l}_{e}(\mathbf{\bar{w}}_l(i),\mathbf{p}_{l}(i)) \label{eq:upperboundser}
\end{equation}
where $\mathcal{\bar{P}}^{l}_{e}(\mathbf{\bar{w}}_l(i),\mathbf{p}_{l}(i))=\mathcal{P}^{l,R}_{e}(\mathbf{\bar{w}}_l(i),\mathbf{p}_{l}(i))+\mathcal{P}^{l,I}_{e}(\mathbf{\bar{w}}_l(i),\mathbf{p}_{l}(i))$.
For small values of SER, since the true SER $\mathcal{P}^{l}_{e}(\mathbf{\bar{w}}_l(i),\mathbf{p}_{l}(i))$ is quite close to the upper bound $\mathcal{\bar{P}}^{l}_{e}(\mathbf{\bar{w}}_l(i),\mathbf{p}_{l}(i))$, the solution obtained by minimizing (\ref{eq:upperboundser}) is practically
relevant and thus we pursue this strategy.

 The output of the reduced-rank receive filter corresponding to the $l$-th branch can be expressed as
\begin{equation}
\begin{split}
x^{l}_0(i)&=\omega^{l}_{0,0}(i)b_0(i)+\overbrace{\sum^{P+L_p-2}_{\mu=1}\omega^{l}_{0,\mu}(i)b_0(i-\mu)}^{{\emph{residual ISI}}}\\&\quad+\overbrace{\sum^{K-1}_{k=1}\sum^{P+L_p-2}_{\mu=0}\omega^{l}_{k, \mu}(i)b_k(i-\mu)}^{{\emph{residual MUI}}}+e(i).\label{eq:qam1}
\end{split}
\end{equation}
 In this expression, we have $\omega^{l}_{k,\mu}(i)=\mathbf{\bar{w}}^{H}_l(i)\mathbf{T}_l\mathbf{P}^{H}_l(i)\mathbf{h}_{k, \mu}(i)$,
  where $\mathbf{h}_{k,\mu}(i)$ denotes the $\mu$-th column vector of matrix $[\mathbf{H}^{T}_{0,k}(i),\ldots,\mathbf{H}^{T}_{L-1,k}(i)]^{T}$, $\mu \in\{0,\ldots,P+L_p-2\}$,  $k\in\{0,\ldots, K-1\}$, while
  the term  $e(i)=\mathbf{\bar{w}}^{H}_l(i)\mathbf{T}_l\mathbf{P}^{H}_l(i)\mathbf{n}(i)$.
Note that the multiplier of the desired symbol, i.e., $\omega^{l}_{0, 0}(i)$ in (\ref{eq:qam1}) is a complex number which can be represented as  $\omega^{l}_{0,0}(i)=\mathbf{\bar{w}}^{H}_l(i)\mathbf{T}_l\mathbf{P}^{H}_l(i)\mathbf{h}_{0,0}(i)=\mathbf{p}^{H}_l(i)\mathbf{D}^{H}(i)\mathbf{h}_{0,0}(i)$.
To simplify the detection  of $M$-QAM symbols, we
apply the following phase rotation operations to transform these complex multipliers into real and positive quantities, which in effect corresponds to the substitution
\begin{equation}
\mathbf{\bar{w}}_l(i)\leftarrow \frac{\omega^{l}_{0, 0}(i)}{|\omega^{l}_{0, 0}(i)|}\mathbf{\bar{w}}_l(i)\label{eq:scalingw}
\end{equation}
or
\begin{equation}
\mathbf{p}_l(i)\leftarrow \frac{\omega^{l}_{0, 0}(i)}{|\omega^{l}_{0, 0}(i)|}\mathbf{p}_l(i)\label{eq:scalingp}.
\end{equation}
By using (\ref{eq:scalingw}) for the reduced-rank receive filter or (\ref{eq:scalingp}) for the preprocessing filter,
we have equivalently $\omega^{l,R}_{0, 0}(i)>0$ and $\omega^{l,I}_{0, 0}(i)=0$, where $\omega^{l,R}_{0, 0}(i)$ and $\omega^{l,I}_{0, 0}(i)$ represent the real and imaginary parts of $\omega^{l}_{0, 0}(i)$ after phase rotation, respectively.
Hence,
the symbol decision $\hat{b}_0(i)=\hat{b}^{R}_0(i)+j\hat{b}^{I}_0(i)$ can be made as (\ref{eq:QAMdecisionreal}) and (\ref{eq:QPSKdecisionimag}),
where $\hat{b}^{R}_0(i)$ and $\hat{b}^{I}_0(i)$ represent the real and imaginary parts of the estimated $M$-QAM symbol, respectively.

We focus on the real part SER, i.e. $\mathcal{P}^{l,R}_{e}(\mathbf{\bar{w}}_l(i),\mathbf{p}_{l}(i))$, to introduce the proposed algorithm; the derivation  regarding the imaginary part SER is straightforward.
  In this case, we have  $N_d=M^{(P+L_p-1)K-1}$ different possible arrangements in total for the MUI and ISI symbols.
 Similar to the BPSK case, for a given $b_0(i)$ we define two sets as follows:
 \setcounter{TempEqCnt}{\value{equation}}
\setcounter{equation}{38}
 \begin{equation}
 \mathcal{\bar{X}}=\{\mathbf{\tilde{b}}^{0}, \mathbf{\tilde{b}}^{1}, \ldots, \mathbf{\tilde{b}}^{N_{d}-1}\}
 \end{equation}
 \begin{equation}
 \mathcal{\bar{Y}}=\{\bar{x}^{l,q}_0=\mathbf{\bar{w}}^{H}_l(i)\mathbf{T}_l\mathbf{P}^{H}_l(i)\mathbf{\tilde{H}}(i)\mathbf{\tilde{b}}^{q}: \ 0\leq q \leq N_d-1 \}.
 \end{equation}
 The conditional PDF of the real part of the reduced-rank receive filter output  for branch $l$ is given by
 \begin{equation}
 \begin{split}
f_l(x|b_{0}(i))&=\frac{1}{ N_d \sigma\sqrt{2\pi\mathbf{\bar{w}}^{H}_l(i)\mathbf{T}_l\mathbf{P}^{H}_l(i)\mathbf{P}_l(i)\mathbf{T}^{H}_l\mathbf{\bar{w}}_l(i)}}\\&\quad\times\sum^{N_d-1}_{q=0}e^{-\frac{|x-\Re[\bar{x}^{l,q}_0]|^2}{2\sigma^2\mathbf{\bar{w}}^{H}_l(i)\mathbf{T}_l\mathbf{P}^{H}_l(i)\mathbf{P}_l(i)\mathbf{T}^{H}_l\mathbf{\bar{w}}_l(i)}}\label{eq:conditionalpdfqam}
\end{split}
\end{equation}
where $\bar{x}^{l,q}_0 \in \mathcal{\bar{Y}}$.
  By using the kernel density estimation, the  single-sample estimate of the PDF similar to (\ref{eq:singlesampleestimatepdf}) can be obtained.
Based on the discussion in \cite{shengchen},
 the  SER expression of the real part for branch $l$ is given by
{\small
 \begin{equation}
 \mathcal{P}^{l,R}_{e}(\mathbf{\bar{w}}_l(i),\mathbf{p}_{l}(i))=\phi\int^{\omega^{l}_{0, 0}(i)(\Re[b_0(i)]-1)}_{-\infty}\frac{1}{\sqrt{2\pi}\rho}e^{-\frac{|x-\Re[{x}^{l}_0(i)]|^2}{2\rho^2}}dx\label{eq:SERfunctionqam1}
 \end{equation}}
 where $\phi=\frac{2\sqrt{M}-2}{\sqrt{M}}$.
 Then, introducing  $s=\frac{x-\Re[{x}^{l}_0(i)]}{\sqrt{2}\rho}$,  (\ref{eq:SERfunctionqam1}) can be rewritten as
 \begin{equation}
\mathcal{P}^{l,R}_{e}(\mathbf{\bar{w}}_l(i),\mathbf{p}_{l}(i))=\phi\int^{\frac{\omega^{l}_{0, 0}(i)(\Re[b_0(i)]-1)-\Re[{x}^{l}_0(i)]}{\sqrt{2}\rho}}_{-\infty}\frac{1}{\sqrt{\pi}}e^{-|s|^2}ds.\label{eq:SERfunctionqam2}
\end{equation}

\setcounter{TempEqCnt}{\value{equation}}
\setcounter{equation}{50}
\begin{figure*}[!t]
\begin{equation}
\bar{\mathbf{w}}^{D_{max}}_l(i)=[\bar{w}_{l, 0}(i), \bar{w}_{l, 1}(i), \ldots, \bar{w}_{l, D_{min}-1}(i), \ldots, \bar{w}_{l, D_{max}-1}(i) ]^{T}\label{eq:Dmax}
\end{equation}
\end{figure*}
\begin{figure*}[!t]
\begin{equation}
{\mathbf{p}}^{I_{max}}_l(i)=[{p}_{l, 0}(i), {p}_{l, 1}(i), \ldots, {p}_{l, I_{min}-1}(i), \ldots, {p}_{l, I_{max}-1}(i) ]^{T}.\label{eq:Imax}
\end{equation}
\end{figure*}
\setcounter{equation}{\value{TempEqCnt}}

\begin{table*}[t]
\centering
 \caption{\normalsize   Adaptive MSER-JPDF reduced-rank algorithm for QAM Symbols} {
\begin{tabular}{ll}
\hline
 $1$ & Set step-size values $\mu_{w}$ and $\mu_{S_{D}}$ and the number of branches $B$.\\
 $2$ & Initialize $\mathbf {\bar{w}}_l(0)$, $\mathbf {p}_{l}(0)$ and $\omega^l_{0, 0}(0)$. Set $\mathbf{T}_{0},\ldots, \mathbf{T}_{B-1}$.\\
 $3$ &   for  time instant $i\in \{0,1,\ldots\}$ do \\
 $4$ & ~ for $l \in \{0, \ldots, B-1\}$ do\\
 $5$ & ~~~  Compute  $\mathbf {p}_{l}(i+1)$ based on (\ref{eq:adaptiveSG2}) using $\omega^{l}_{0, 0}(i)$, $\mathbf{T}_{l}$ and $\mathbf{\bar{w}}_l(i)$. \\
 $6$ & ~~~ Compute $\omega^{l}_{0, 0}(i)$ based on $\omega^{l}_{0, 0}(i)=\mathbf{p}^{H}_l(i+1)\mathbf{D}^{H}(i)\mathbf{h}_{0,0}(i)$. \\
     & ~~~ Adjust $\mathbf{p}_l(i+1)$ by using $\mathbf{p}_l(i+1)\leftarrow \frac{\omega^{l}_{0, 0}(i)}{|\omega^{l}_{0, 0}(i)|}\mathbf{p}_l(i+1)$. \\
  $7$ & ~~~ Compute $\mathbf{\bar{w}}_l(i+1)$ based on  (\ref{eq:adaptiveSG1}) using  $\omega^{l}_{0, 0}(i)$, $\mathbf{T}_{l}$ and $\mathbf{p}_l(i+1)$. \\
 $8$ & ~~~  Compute $\omega^{l}_{0, 0}(i+1)$ based on $\omega^{l}_{0, 0}(i+1)=\mathbf{\bar{w}}^{H}_l(i+1)\mathbf{T}_l\mathbf{P}^{H}_l(i+1)\mathbf{h}_{0,0}(i)$.\\
  & ~~~  Adjust $\mathbf{\bar{w}}_l(i+1)$ by using $\mathbf{\bar{w}}_l(i+1)\leftarrow \frac{\omega^{l}_{0, 0}(i+1)}{|\omega^{l}_{0, 0}(i+1)|}\mathbf{\bar{w}}_l(i+1)$.\\
 $9$ & ~ end  \\
 $10$ & ~ Select the optimal branch $l_{opt}=\arg \min_{0\leq l \leq B-1} |b_0(i)-{x}^{l}_{0}(i)|$, where ${x}^{l}_{0}(i)=\mathbf{\bar{w}}^{H}_l(i+1)\mathbf{T}_{l}\mathbf{P}^{H}_{l}(i+1)\mathbf{r}(i)$. \\
     & ~ Generate the estimate symbol corresponding to branch $l_{opt}$: $\hat{b}_0(i)=\mathcal{Q}\{{x}^{l_{opt}}_{0}(i)\}$.\\
\hline
\label{tab:table2}
\end{tabular}
}
\end{table*}

Note that $\omega^{l}_{0, 0}(i)$ and ${x}^{l}_0(i)$ are both functions of $\mathbf{\bar{w}}_l(i)$ and $\mathbf{p}_l(i)$. By taking the gradient of (\ref{eq:SERfunctionqam2})
with respect to $\mathbf {\bar{w}}^{*}_l(i)$ and $\mathbf{p}^{*}_l(i)$, respectively,
we obtain
{\small
\begin{equation}
\begin{split}
\frac{\partial \mathcal{P}^{l,R}_{e}(\mathbf{\bar{w}}_l(i),\mathbf{p}_{l}(i))}{\partial \mathbf{\bar{w}}^{*}_l} &
=\frac{\phi}{\sqrt{2\pi}\rho}e^{- \frac{|\omega^{l}_{0, 0}(i)(\Re[b_0(i)]-1)-\Re[{x}^{l}_0(i)]|^2}{2\rho^2}}\\&\quad\times\mathbf{T}_l\mathbf{P}^{H}_l(i)\big((\Re[b_0(i)]-1)\mathbf{h}_{0,0}(i)-\mathbf{r}(i)\big)
\label{eq:SGQAM1}
\end{split}
\end{equation}
}
and
 {\small \begin{equation}
 \begin{split}
 \frac{\partial \mathcal{P}^{l,R}_{e}(\mathbf{\bar{w}}_l(i),\mathbf{p}_{l}(i))}{\partial \mathbf{p}^{*}_l} &
=\frac{\phi}{\sqrt{2\pi}\rho}e^{- \frac{|\omega^{l}_{0, 0}(i)(\Re[b_0(i)]-1)-\Re[{x}^{l}_0(i)]|^2}{2\rho^2}}\\&\quad\times\mathbf{D}^{H}(i)\big((\Re[b_0(i)]-1)\mathbf{h}_{0,0}(i)-\mathbf{r}(i)\big).
\label{eq:SGQAM2}
\end{split}
 \end{equation}
 }
 Following a similar approach, we can obtain the  SER expression of the imaginary part corresponding to the $l$-th branch as
{\small
 \begin{equation}
\mathcal{P}^{l,I}_{e}(\mathbf{\bar{w}}_l(i),\mathbf{p}_{l}(i))=\phi\int^{\omega^{l}_{0, 0}(i)(\Im[b_0(i)]-1)}_{-\infty}\frac{1}{\sqrt{2\pi}\rho}e^{-\frac{|x-\Im[{x}^{l}_0(i)]|^2}{2\rho^2}}dx. \label{eq:SERfunctionqam23}
\end{equation}
}
 The gradient terms of this expression with respect $\mathbf{\bar{w}}_l(i)$ and $\mathbf{p}_l(i)$ are given by
{\small
\begin{equation}
\begin{split}
\frac{\partial \mathcal{P}^{l,I}_{e}(\mathbf{\bar{w}}_l(i),\mathbf{p}_{l}(i))}{\partial \mathbf{\bar{w}}^{*}_l} &
=\frac{\phi}{\sqrt{2\pi}\rho}e^{- \frac{|\omega^{l}_{0, 0}(i)(\Im[b_0(i)]-1)-\Im[{x}^{l}_0(i)]|^2}{2\rho^2}}\\&\quad\times\mathbf{T}_l\mathbf{P}^{H}_l(i)\big(j \mathbf{r}(i)+(\Im[b_0(i)]-1)\mathbf{h}_{0,0}(i)\big)
\label{eq:SGQAM3}
\end{split}
\end{equation}
}
and
{\small
 \begin{equation}
 \begin{split}
 \frac{\partial \mathcal{P}^{l,I}_{e}(\mathbf{\bar{w}}_l(i),\mathbf{p}_{l}(i))}{\partial \mathbf{p}^{*}_l} &
=\frac{\phi}{\sqrt{2\pi}\rho}e^{- \frac{|\omega^{l}_{0, 0}(i)(\Im[b_0(i)]-1)-\Im[{x}^{l}_0(i)]|^2}{2\rho^2}}\\&\quad\times\mathbf{D}^{H}(i)\big(j\mathbf{r}(i)+(\Im[b_0(i)]-1)\mathbf{h}_{0,0}(i)\big)
\label{eq:SGQAM4}.
\end{split}
 \end{equation}
 }

Next, the preprocessing filter and the reduced-rank
receive filter  are jointly optimized using recursions that are employed in an alternating fashion.
The proposed SG update equations corresponding to branch $l$ for QAM symbols can be
obtained by substituting the gradient terms (\ref{eq:SGQAM1}), (\ref{eq:SGQAM2}), (\ref{eq:SGQAM3}) and
(\ref{eq:SGQAM4}) in the following expressions:
\begin{equation}
\begin{split}
\mathbf{\bar{w}}_l(i+1)&=\mathbf{ \bar{w}}_l(i)-\mu_{w}\bigg(\frac{\partial \mathcal{P}^{l,R}_{e}(\mathbf{\bar{w}}_l(i),\mathbf{p}_{l}(i))}{\partial \mathbf{\bar{w}}^{*}_l}\\&\quad+ \frac{\partial \mathcal{P}^{l,I}_{e}(\mathbf{\bar{w}}_l(i),\mathbf{p}_{l}(i))}{\partial \mathbf{\bar{w}}^{*}_l}\bigg)\label{eq:adaptiveSG1}
\end{split}
\end{equation}
and
\begin{equation}
\begin{split}
\mathbf{p}_{l}(i+1)&=\mathbf{p}_{l}(i)-\mu_{p}\bigg( \frac{\partial \mathcal{P}^{l,R}_{e}(\mathbf{\bar{w}}_l(i),\mathbf{p}_{l}(i))}{\partial \mathbf{p}^{*}_l}\\&\quad+ \frac{\partial \mathcal{P}^{l,I}_{e}(\mathbf{\bar{w}}_l(i),\mathbf{p}_{l}(i))}{\partial \mathbf{p}^{*}_l}\bigg).\label{eq:adaptiveSG2}
\end{split}
\end{equation}
Similar to the proposed adaptive algorithm for BPSK symbols, in this case
the algorithm is developed to start in the  TR mode, and then to switch to
the DD mode.
The
proposed MSER-JPDF algorithm for  $M$-QAM symbols is summarized in Table
\ref{tab:table2}\footnote{Note that the proposed adaptive algorithms can be extended to multicarrier systems (e.g. filter-bank multicarrier  and generalized orthogonal frequency-division multiplexing (OFDM) systems that are being considered for $5$G).  With multicarrier modulation,  the system can be described with a flat fading channel for each subcarrier and the proposed scheme and algorithms would mitigate the MUI rather than both ISI and MUI, as with the current description. }, where the arrow denotes an overwrite operation.

\subsection{Automatic Parameter Selection}

\begin{table*}[t]
\centering%
\caption{\normalsize Computational complexity in the case of BPSK} {
\begin{tabular}{ccc}
\hline \rule{0cm}{2.5ex}&  \multicolumn{2}{c}{Number of operations
per  symbol} \\ \cline{2-3}
Algorithm & {Multiplications} & {Additions} \\
\hline
{\small  Full-Rank-LMS}  & {\small $2LP+1$} & {\small $2LP$}  \\
{\small  Full-Rank-MSER} \cite{shengchensp2008}  &  {\small $3LP+1$} & {\small $2LP$}  \\
{\small  MSER-JIO } \cite{caicl2013} &  {\small $8LPD +7D+2LP + 9$} &  {\small $7LPD+2LP-1$}  \\
{\small   EIG} \cite{eign2} &   {\small $O((LP)^3)$} &  {\small $O((LP)^3)$}  \\
{\small  MSER-MSWF} \cite{mber3} &  {\small $D(LP)^2+4LPD+5D+LP+7$} &  {\small $D(LP)^2+5LPD-1$} \\
{\small  MSER-JPDF} & {\small $BI(LP+1.5)+BD(I+2)+6B-0.5I^2B$}  & {\small $BI(LP+0.5)+BD(I+1)-0.5I^2B$} \\
\hline
\label{tab:table3}
\end{tabular}
}
\end{table*}

\begin{table*}[t]
\centering%
\caption{\normalsize Computational complexity in the case of QAM} {
\begin{tabular}{ccc}
\hline \rule{0cm}{2.5ex}&  \multicolumn{2}{c}{Number of operations
per  symbol} \\ \cline{2-3}
Algorithm & {Multiplications} & {Additions} \\
\hline
{\small  Full-Rank-LMS}  & {\small $2LP+1$} & {\small $2LP$}  \\
{\small  Full-Rank-MSER} \cite{shengchen}  &  {\small $6LP+5$} & {\small $5LP$}  \\
{\small  MSER-JIO } \cite{caicl2013} &  {\small $10LPD+7D+4LP+17$} &  {\small $9LPD+4LP+3$}  \\
{\small   EIG} \cite{eign2} &   {\small $O((LP)^3)$} &  {\small $O((LP)^3)$}  \\
{\small  MSER-MSWF} \cite{mber3} &  {\small $D(LP)^2+5LPD+5D+2LP+11$} &  {\small $D(LP)^2+6LPD+LP+1$} \\
{\small  MSER-JPDF} & {\small $BI(2LP+3)+2BD(2I+1)+8B-BI^2$}  & {\small $2BI(LP+1)+BD(4I-1)+2BLP-BI^2$} \\
\hline
\label{tab:table4}
\end{tabular}
}
\end{table*}

Since the performance of the adaptive  MSER-JPDF reduced-rank algorithm depends on the rank $D$ and the length of the preprocessor $I$, we develop an  automatic parameter selection scheme in which $D$ and $I$ are adjusted on-line for added flexibility in the implementation of the proposed structure.
 The proposed scheme performs the search within a range of appropriate values and relies on the  Euclidean distance to determine the lengths of $\mathbf{\bar{w}}_l(i)$ and $\mathbf{p}_l(i)$ corresponding to branch $l$, which can be adjusted in a flexible structure. At each time instant, for branch $l$ we adapt a reduced-rank receive filter $\bar{\mathbf{w}}^{D_{max}}_l(i)$ and a preprocessor ${\mathbf{p}}^{I_{max}}_l(i)$ according to the proposed algorithms in Table \ref{tab:table1} or \ref{tab:table2} with the maximum allowed rank $D_{max}$ and the maximum preprocessor length $I_{max}$, respectively, which can be expressed as (\ref{eq:Dmax}) and (\ref{eq:Imax}).
Then, we test the values of rank $D$ and preprocessor length $I$ within the range, namely, $D_{min} \leq D \leq D_{max}$ and $I_{min} \leq I \leq I_{max}$. For each pair of $D$ and $I$ of branch $l$, we substitute the vectors $\bar{\mathbf{w}}^{(D)}_{l}(i)=[\bar{w}_{l, 0}(i), \ldots, \bar{w}_{l, D-1}(i)]^{T}$ and ${\mathbf{p}}^{(I)}_{l}(i)=[{p}_{l, 0}(i), \ldots, {p}_{l, I-1}(i)]^{T}$, whose components are taken from
$\bar{\mathbf{w}}^{D_{max}}_l(i)$ and ${\mathbf{p}}^{I_{max}}_l(i)$ in (\ref{eq:Dmax}) and (\ref{eq:Imax}), into
 the following expression of the Euclidean distance
 \setcounter{TempEqCnt}{\value{equation}}
\setcounter{equation}{52}
\begin{equation}
{\varepsilon}^{D, I}_l(i)=|b_0(i)-\bar{\mathbf{w}}^{(D) H}_{l}(i)\mathbf{T}_{l}\mathbf{R}(i){\mathbf{p}}^{(I)*}_{l}(i)|.\label{eq:varepmax}
\end{equation}
The optimum lengths $D_{l, opt}$ and $I_{l, opt}$ for the reduced-rank receive filter and the preprocessor corresponding to branch $l$ at time instant $i$ can be chosen as follows:
\begin{equation}
[D_{l, opt}, I_{l, opt}]=\arg \min_{\substack{D_{min}\leq D\leq D_{max} \\ I_{min}\leq I\leq I_{max}}} {\varepsilon}^{D, I}_l(i).\label{eq:searchImaxDmax}
\end{equation}
After the optimum filter lengths are determined for all the branches, we select the optimum branch of the JPDF scheme  based on the following criterion
\begin{equation}
l_{opt}=\arg \min_{0\leq l \leq B-1} |b_0(i)-\bar{\mathbf{w}}^{(D_{l,opt}) H}_l(i)\mathbf{T}_l\mathbf{R}(i){\mathbf{p}}^{(I_{l,opt})*}_l(i)|.
\end{equation}
Hence, in this case the output of the proposed reduced-rank scheme at time instant $i$ is generated according to the selected  branch $l_{opt}$ with the optimum parameters.
 For given lengths $D_{max}$ and $I_{max}$, the complexity of the automatic parameter selection scheme mainly lies in the computation of the distance ${\varepsilon}^{D, I}_l(i)$ in (\ref{eq:varepmax})
and a simple search over the candidates $D$ and $I$ in (\ref{eq:searchImaxDmax}).
Consequently, this approach does not considerably increase the computational complexity.
Note that the smaller values of $D$ and $I$ may produce faster
adaptation during the initial stage of the estimation procedure
while  slightly larger values of $D$ and $I$ usually yield  better steady-state
performance.
In Section \ref{Section6:simulations}, we will show that the proposed adaptive MSER-JPDF reduced-rank algorithms with the automatic parameter scheme can improve the convergence speed and steady-state performance compared to the algorithms with  fixed parameters  $D$ and $I$.

\section{Analysis of the Proposed Algorithms}
\label{Section5:analysis}

In this section, we carry out the analysis of the  proposed MSER-JPDF adaptive reduced-rank algorithms.
Firstly, we compare the computational complexity  of these new algorithms to that of  existing reduced-rank adaptive algorithms and of the conventional full-rank adaptive algorithms.
Secondly, we investigate the convergence behavior of the newly proposed algorithms.

\subsection{Computational Complexity}

\begin{figure}[!hhh]
\centering
\scalebox{0.42}{\includegraphics{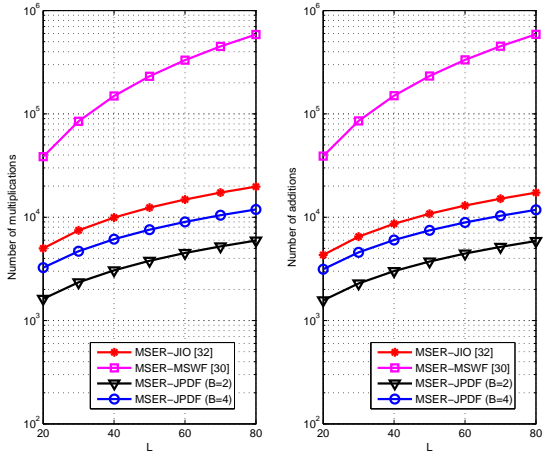}}
\caption{ Comparison of computational  complexity for the recently reported reduced-rank algorithms and the proposed MSER-JPDF reduced-rank algorithms in the case of BPSK symbols ($I=12$, $D=10$, $B=2,4$).}
\label{fig:overallcomplexity1}
\end{figure}

\begin{figure}[!hhh]
\centering
\scalebox{0.42}{\includegraphics{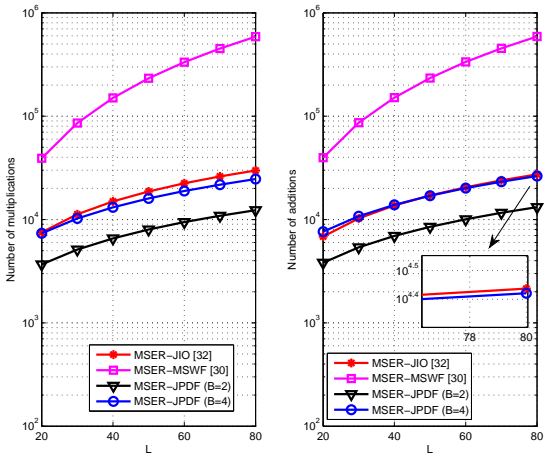}}
\caption{ Comparison of computational complexity for the recently reported reduced-rank algorithms and the proposed MSER-JPDF reduced-rank algorithms in the case of QAM symbols ($I=12$, $D=10$, $B=2,4$).}
\label{fig:overallcomplexity}
\end{figure}

In Table \ref{tab:table3}, we focus on the case of BPSK and show the number of additions and
multiplications per symbol of the proposed adaptive  reduced-rank algorithm, the existing
adaptive MSER-based reduced-rank algorithms \cite{mber3}, \cite{caicl2013}, the conventional adaptive
LMS full-rank algorithm \cite{haykin} and the adaptive full-rank algorithm based on the SER
criterion \cite{shengchensp2008}.
Table \ref{tab:table4} shows the corresponding figures for the case of
QAM symbols.
The overall complexity of the proposed
algorithm includes the complexity of the selection mechanism and the design complexity of each branch  multiplied by the number of branches  $B$.
In practice, to limit computational complexity,  the number of branches should be kept small, typically $B \in \{2,3,4\}$ in this work.
 From the figures in these Tables, it is clear that the full-rank SG-based algorithms have lower complexity than all the reduced-rank algorithms under comparison. As pointed out earlier, the main problem of the full-rank SG-based adaptive filtering algorithms is their very slow convergence performance when a filter with a large number of adaptive weights is employed, as needed here to process a large-dimensional received data vectors.
Although the complexity of the reduced-rank algorithms is higher than that of the full-rank SG-based algorithms, in Section \ref{Section6:simulations} we will show that for a limited increase in complexity, they can yield a significant increase in convergence speed and tracking performance when compared to the former, whose performance in the current application is simply unacceptable.
In Fig. \ref{fig:overallcomplexity1} and Fig. \ref{fig:overallcomplexity}, we show the  computational complexity against the number of received antenna elements for the recently reported MSER-based reduced-rank algorithms \cite{mber3}, \cite{caicl2013} and the proposed MSER-JPDF reduced-rank algorithms.
In particular, we consider a commonly used configuration with $P=3$, $I=12$ and $D=10$.
From these results, we find that  the proposed algorithms have significantly lower complexity than the existing MSER-MSWF reduced-rank algorithm\footnote{Note that the MSER-based MSWF reduced-rank algorithm \cite{mber3} corresponds to the use of the procedure in \cite{MWF} to construct $\mathbf{S}_D(i)$ and of the MSER-SG adaptive algorithm in \cite{shengchensp2008}, \cite{shengchen}, to compute $\mathbf{ \bar{w}}(i)$.} \cite{mber3}. Although the complexity of the proposed algorithms increases as the number $B$ of branches increases,  it remains lower than that of the MSER-JIO reduced-rank algorithms \cite{caicl2013} for a large number of antenna elements.
 Note that for the proposed adaptive MSER-JPDF reduced-rank algorithms with the automatic selection scheme, the numbers of multiplications and additions are the same as those shown in Table \ref{tab:table3} and \ref{tab:table4}, but with parameters $D_{max}$ and $I_{max}$ instead of $D$ and $I$. In addition, a simple search over the values of ${\varepsilon}^{D, I}_l(i)$ is performed in order to select the optimal parameter values  $D_{l, opt}$ and $I_{l, opt}$, whose time complexity  is about $1.5(D_{max}-D_{min})(I_{max}-I_{min})$.
In Section \ref{Section6:simulations}, we will show that  the estimation and detection  performance of the proposed  MSER-JPDF exceeds that of  existing adaptive reduced-rank algorithms by a wide margin.

\subsection{Convergence Analysis}

In the following, we discuss the convergence behavior of the proposed algorithms for the joint design of the preprocessing filter and reduced-rank receive filter. Firstly, we focus on the adaptive MSER-JPDF reduced-rank algorithm
 with BPSK symbols, and provide a proof for its convergence by considering a  given branch, i.e., a fixed value of $l\in\{0, 1, \ldots, B-1\}$.
From  \cite{shengchensp2008}, \cite{shengchen}, we can see that for the  average SER cost function $\mathcal{\tilde{P}}^{l}_{e}(\mathbf{\bar{w}}_l,\mathbf{p}_l)$ corresponding to the $l$-th branch, by fixing the preprocessor  $\mathbf{p}_{l}$, there will exist infinitely many global MSER solutions for the reduced-rank receive filter $\mathbf{\bar{w}}_l$, which form an infinite half line  in the filter weight space\footnote{By fixing the preprocessing filter, based on the discussion in \cite{shengchensp2008}-\cite{xfwangtsp2000}, we obtain that any local minimizer of the SER cost function is a global minimizer. Let $\mathbf{\bar{w}}_{l,opt}$ be such a global MSER solution for the $l$-th branch. According to  \cite{shengchensp2008} the
weight vectors $a\mathbf{\bar{w}}_{l,opt}$, $a>0$, are all global MSER
solutions for branch $l$, which form an infinite half line in the
filter weight space.}. Since the cost function is symmetric in  $\mathbf{\bar{w}}_l$ and $\mathbf{p}_l$, similarly by fixing $\mathbf{\bar{w}}_l$, there will be multiple global MSER solutions for $\mathbf{p}_l$. In the proposed SG-based adaptive algorithm, for each branch $l$ at time index $i$, we try to find the   gradient direction along the instantaneous SER cost function surface for $\mathbf{p}_l(i)$ and $\mathbf{\bar{w}}_l(i)$, respectively, in order to compute $\mathbf{p}_l(i+1)$ and $\mathbf{\bar{w}}_l(i+1)$.
By using small step sizes,  $N$ SG iterative steps are approximately equivalent, on average, to a single larger  step in the direction of steepest decent on the average SER performance surface $\mathcal{\tilde{P}}^{l}_{e}(\mathbf{\bar{w}}_l,\mathbf{p}_l)$, where $N$ is a large integer \cite{haykin}.
Hence, the instantaneous gradient can be replaced by a less noisy average, leading to  the following steepest descent update expressions:
\begin{equation}
\mathbf{p}_l((m+1)N)=\mathbf{p}_l(mN)-\alpha_p\frac{\mathcal{\tilde{P}}^{l}_{e}(\mathbf{\bar{w}}_l(mN),\mathbf{p}_{l}(mN))}{\partial \mathbf{p}^{*}_l}\label{eq:steepest1}
\end{equation}
\begin{equation}
\mathbf{w}_l((m+1)N)=\mathbf{w}_l(mN)-\alpha_w\frac{\mathcal{\tilde{P}}^{l}_{e}(\mathbf{\bar{w}}_l(mN),\mathbf{p}_{l}((m+1)N))}{\partial \mathbf{\bar{w}}^{*}_l}\label{eq:steepest2}
\end{equation}
where $m\in\{0, 1, 2, \ldots\}$ denotes the index of blocks of received vectors, and $\alpha_w$ and $\alpha_p$ are the step sizes for the reduced-rank receive and  preprocessing filters, respectively.
Therefore, by following (\ref{eq:steepest1}) and (\ref{eq:steepest2}) we can obtain the following inequalities for the average SER corresponding to the $l$-th branch:
\begin{equation}
\mathcal{\tilde{P}}^{l}_{e}(\mathbf{p}_{l}((m+1)N),\mathbf{\bar{w}}_{l}(mN))\leq\mathcal{\tilde{P}}^{l}_{e}(\mathbf{p}_{l}(mN),\mathbf{\bar{w}}_{l}(mN))\label{eq:analysis1}
\end{equation}
and
{\small
\begin{equation}
\mathcal{\tilde{P}}^{l}_{e}(\mathbf{p}_{l}((m+1)N),\mathbf{\bar{w}}_{l}((m+1)N))\leq\mathcal{\tilde{P}}^{l}_{e}(\mathbf{p}_{l}((m+1)N),\mathbf{\bar{w}}_{l}(mN)).\label{eq:analysis2}
\end{equation}
}
This algorithm starts from the initial values $\mathbf{\bar{w}}_l(0)$ and $\mathbf{p}_l(0)$. By using (\ref{eq:analysis1}) and (\ref{eq:analysis2}),   we obtain
 \begin{equation}
\begin{split}
\ldots&\leq\mathcal{\tilde{P}}^{l}_{e}(\mathbf{p}_{l}((m+1)N),\mathbf{\bar{w}}_{l}((m+1)N))\\&\leq \mathcal{\tilde{P}}^{l}_{e}(\mathbf{p}_{l}((m+1)N),\mathbf{\bar{w}}_{l}(mN))\\&\leq \mathcal{\tilde{P}}^{l}_{e}(\mathbf{p}_{l}(mN),\mathbf{\bar{w}}_{l}(mN))\leq
\ldots \leq \mathcal{\tilde{P}}^{l}_{e}(\mathbf{p}_{l}(N),\mathbf{\bar{w}}_{l}(N))
\\&\leq\mathcal{\tilde{P}}^{l}_{e}(\mathbf{p}_{l}(N),\mathbf{\bar{w}}_{l}(0))\leq \mathcal{\tilde{P}}^{l}_{e}(\mathbf{p}_{l}(0),\mathbf{\bar{w}}_{l}(0)).
\label{eq:analysis6}
\end{split}
\end{equation}
This shows that as the number of iterations increases the average SER  of each branch $l$ decreases. Therefore, since the SER value is lower bounded by $0$, the adaptive MSER-JPDF reduced-rank algorithm for each branch is convergent.
 Furthermore, at each time index after updating the reduced-rank receive and  preprocessing filters  for all the branches, we select the optimal branch $l_{opt}$  corresponding to the minimum Euclidean distance between the true transmit symbol and the filter output of each branch (see (\ref{eq:errorpl})). By taking the average over independent realizations, we obtain that the average SER of the proposed multiple filtering branches scheme  is lower than  the average SER of each branch and it  decreases with  increasing of the number of iterations.

Compared to the proposed algorithm for BPSK symbols in Table \ref{tab:table1}, the proposed algorithm for $M$-QAM symbols in Table \ref{tab:table2} has two extra operations, namely steps $6$ and $8$. Those operations are carried out after each adaptive iteration to adjust the preprocessing  and the reduced-rank receive filters, respectively, in order to guarantee that $\omega_{0, 0}$ is positive real as needed for $M$-QAM detection; however, these operations do not affect the convergence.  Thus,
for each branch  we still have $\mathcal{\tilde{P}}^{l}_e(\mathbf{p}_l((m+1)N),\mathbf{\bar{w}}_l((m+1)N))\leq\mathcal{\tilde{P}}^{l}_e(\mathbf{p}_l((m+1)N),\mathbf{\bar{w}}_l(mN))\leq \mathcal{\tilde{P}}^{l}_e(\mathbf{p}_l(mN),\mathbf{\bar{w}}_l(mN))$ at the $m$-th block of received vectors. Hence, a similar analysis as above  can be done  for the adaptive MSER-JPDF reduced-rank algorithm with $M$-QAM symbols, thereby showing the convergence of this algorithm.

\section{Simulations}
\label{Section6:simulations}

In this section, we evaluate the  performance of the proposed adaptive
MSER-JPDF reduced-rank algorithms and compare them with existing adaptive
full-rank and reduced-rank algorithms. Monte-carlo simulations are
conducted to verify the effectiveness of the MSER-JPDF adaptive
reduced-rank SG algorithms for large-scale multiple-antenna systems.
In the simulations, we assume that the base station is equipped with $L=40$ antenna elements and the system includes one desired user and $K-1$ interfering users. We adopt an observation window of $P=3$ symbols, the multipath channels (the channel vectors $[h_{k, \nu, 0}(i),\ldots, h_{k, \nu, L_p-1}(i)]^{T}$) are modeled by FIR filters, with the $L_p$ coefficients spaced by one symbol. The channel are time varying over the transmitted symbols, where the profile follows the Universal Mobile Telecommunications System (UMTS) Vehicular A channel model \cite{umtsa} with $L_p=3$, and the fading is given by Clarke's model \cite{rappaport}.
 The  $K$ users have identical transmit power, which is normalized to unity for convenience.
 We set $\rho=1.06\sigma$ \cite{mber2}, \cite{densityestimation1}, \cite{parzene1962} in (\ref{eq:adaptiveSGbpsk1}), (\ref{eq:adaptiveSGbpsk2}),  (\ref{eq:adaptiveSG1}) and (\ref{eq:adaptiveSG2}).
%
%
The
 full-rank,
 reduced-rank  and  preprocessing filters are  initialized as $\mathbf{h}_{0,0}(0)$, $\mathbf{T}_{l}\mathbf{h}_{0,0}(0)$ and $[1,0,\ldots,0]^{T}$ with appropriate dimensions, respectively\footnote{The full-rank and reduced-rank receive filters are initialized based on the desired symbol's matched filter, i.e., $\mathbf{h}_{0,0}(0)$. In this work, we assume that this information is known as a priori knowledge  for simplicity. In practice, it should be obtained by implementing channel estimation algorithms.}.
The algorithms process $300$ symbols in the training mode, which is followed by a decision directed mode of operation.

\begin{figure}[!hhh]
\centering
\scalebox{0.45}{\includegraphics{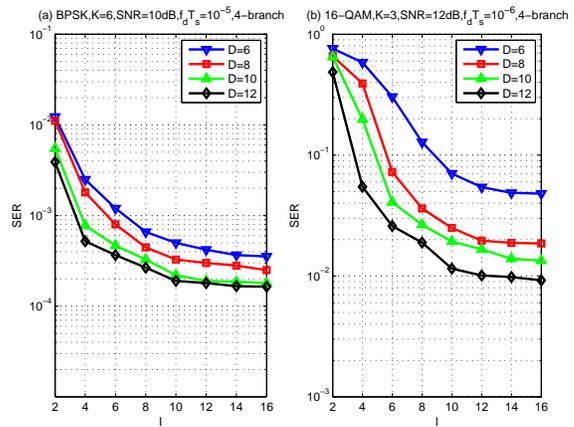}}
\caption{  SER performance versus parameter $I$ with different values of $D$ for the adaptive MSER-JPDF reduced-rank algorithms: (a) BPSK symbols ($K=6$,  SNR $=10$dB, $f_dT_s=10^{-5}$); (b) $16$-QAM symbols ($K=3$,  SNR $=12$dB, $f_dT_s=10^{-6}$).}
\label{fig:simulation1}
\end{figure}

\begin{figure}[!hhh]
\centering
\scalebox{0.45}{\includegraphics{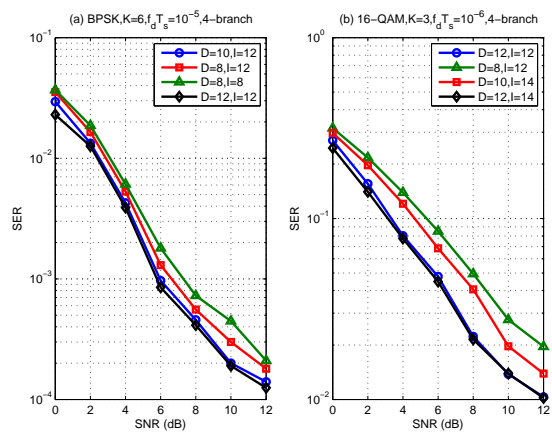}}
\caption{  SER performance of the adaptive MSER-JPDF reduced-rank algorithms versus SNR for   different values of $D$ and $I$: (a) BPSK symbols ($K=6$,  $f_dT_s=10^{-5}$); (b) $16$-QAM symbols ($K=3$,   $f_dT_s=10^{-6}$).}
\label{fig:simulation2}
\end{figure}

In the first experiment, we investigate the effects of $D$ and $I$ on the proposed adaptive  MSER-JPDF reduced-rank algorithm for BPSK symbols and $16$-QAM symbols, respectively. In particular, we consider the case with $B=4$ branches.  In Fig. \ref{fig:simulation1}(a) and (b), we show the  SER performance of the proposed algorithms  (Table \ref{tab:table1} and \ref{tab:table2}) versus $I$ for $D=6, 8, 10, 12$. The SER is evaluated for data records of  $1500$ symbols.
Note that Fig. \ref{fig:simulation1}(a) and (b) correspond to the algorithm with BPSK and $16$-QAM symbols, respectively.
In order to provide the best performance, we tuned $\mu_w=\mu_p=0.01$ for Fig. \ref{fig:simulation1}(a) and $\mu_w=\mu_p=0.006$ for Fig. \ref{fig:simulation1}(b).
For the case with BPSK symbols, we use $10$dB for the input SNR while the normalized Doppler
frequency is $f_dT_s=10^{-5}$,  where $f_d$ is the Doppler frequency in Hz and $T_s$ is the symb\emph{}ol duration.
For the case with $16$-QAM symbols, we set SNR$=12$dB and $f_dT_s=10^{-6}$.
Notice that we have conducted experiments to obtain the most adequate  values of parameters $D$ and $I$ for these algorithms and values beyond the range shown here need not be considered since they do not lead to performance improvements.
From the results of Fig. \ref{fig:simulation1}(a), we can see that the SER initially  decreases with an increase of $I$, but beyond $I=12$ the performance does not change significantly.
In order to keep a low complexity  we adopt  $I=12$ and $D=10$  for the proposed algorithm with fixed $D$ and $I$ in the case of BPSK symbols,
 but also later compare this choice with the proposed algorithm that uses the automatic parameter selection.
 Similarly, based on the results shown in Fig. \ref{fig:simulation1}(b)   we select $D=I=12$ for the proposed algorithm with $16$-QAM symbols.
Fig. \ref{fig:simulation2}(a) and (b) illustrate the  SER performance of the adaptive MSER-JPDF reduced-rank algorithms  versus SNR for   different values of $D$ and $I$, corresponding to the case with BPSK and $16$-QAM symbols, respectively.
We employ data records of $1500$ symbols.
The algorithms are run with $B=4$ branches and
the normalized Doppler
frequency is set to $f_dT_s=10^{-6}$.
From the results in this figure, we can see that the selected values of $D$ and $I$ offer a good tradeoff  between complexity and performance for the proposed algorithms with fixed $D$ and $I$.



\begin{figure}[!hhh]
\centering
\scalebox{0.53}{\includegraphics{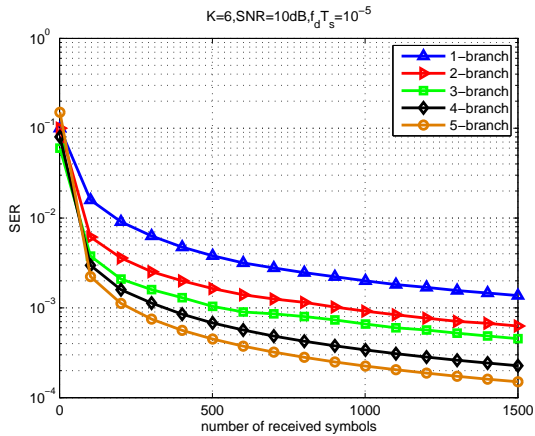}}
\caption{ SER performance of the adaptive MSER-JPDF reduced-rank algorithm with BPSK symbols
versus the number of received symbols for different number of branches ($K=6$,
  SNR $=10$dB,  $f_dT_s=10^{-5}$).}
\label{fig:simulation3}
\end{figure}

\begin{figure}[!hhh]
\centering
\scalebox{0.53}{\includegraphics{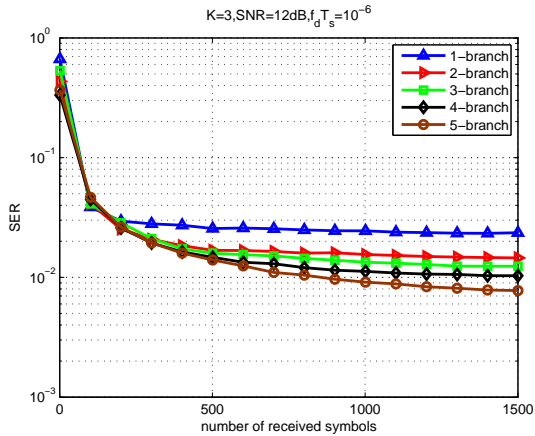}}
\caption{ SER performance of the adaptive MSER-JPDF reduced-rank algorithm with $16$-QAM symbols
versus the number of received symbols for different number of branches ($K=3$,
  SNR $=12$dB,  $f_dT_s=10^{-6}$).}
\label{fig:simulation4}
\end{figure}

In the next experiment, we study the impact of the number
of branches on the performance of the proposed algorithms.
In Fig. \ref{fig:simulation3} and  Fig. \ref{fig:simulation4}, we study  the convergence behavior
of the proposed algorithms by showing the evolution of
the SER performance as a function of the number of received symbols\footnote{The on-line adaptive algorithms update the filters once for each received symbol. The $i$-th symbol time instant corresponds to the $i$-th adaptive iteration, and $T_s$ also corresponds to the time between iterations.}.
 We designed the adaptive MSER-JPDF reduced-rank algorithms with $B=1,2,3,4,5$ parallel branches.
 Fig. \ref{fig:simulation3} shows the performance of the BPSK case, where the other system parameters are set as follows: $K=6$, SNR $=10$dB and $f_dT_s=10^{-5}$; while
 Fig. \ref{fig:simulation4} shows the performance for the $16$-QAM case, where the other parameters are: $K=3$,
  SNR $=12$dB and $f_dT_s=10^{-6}$.
 The values of step-size are adjusted as  in the previous experiment.
 From these results, it can be noted  that the
performance of the proposed MSER-JPDF algorithms improves as the number of
parallel branches, i.e. parameter $B$, increases.
In this regard,
we adopt $B=2$ and $B=4$ for the remaining experiments because they present an interesting
tradeoff between performance and complexity.


%

\begin{figure}[!hhh]
\centering
\scalebox{0.53}{\includegraphics{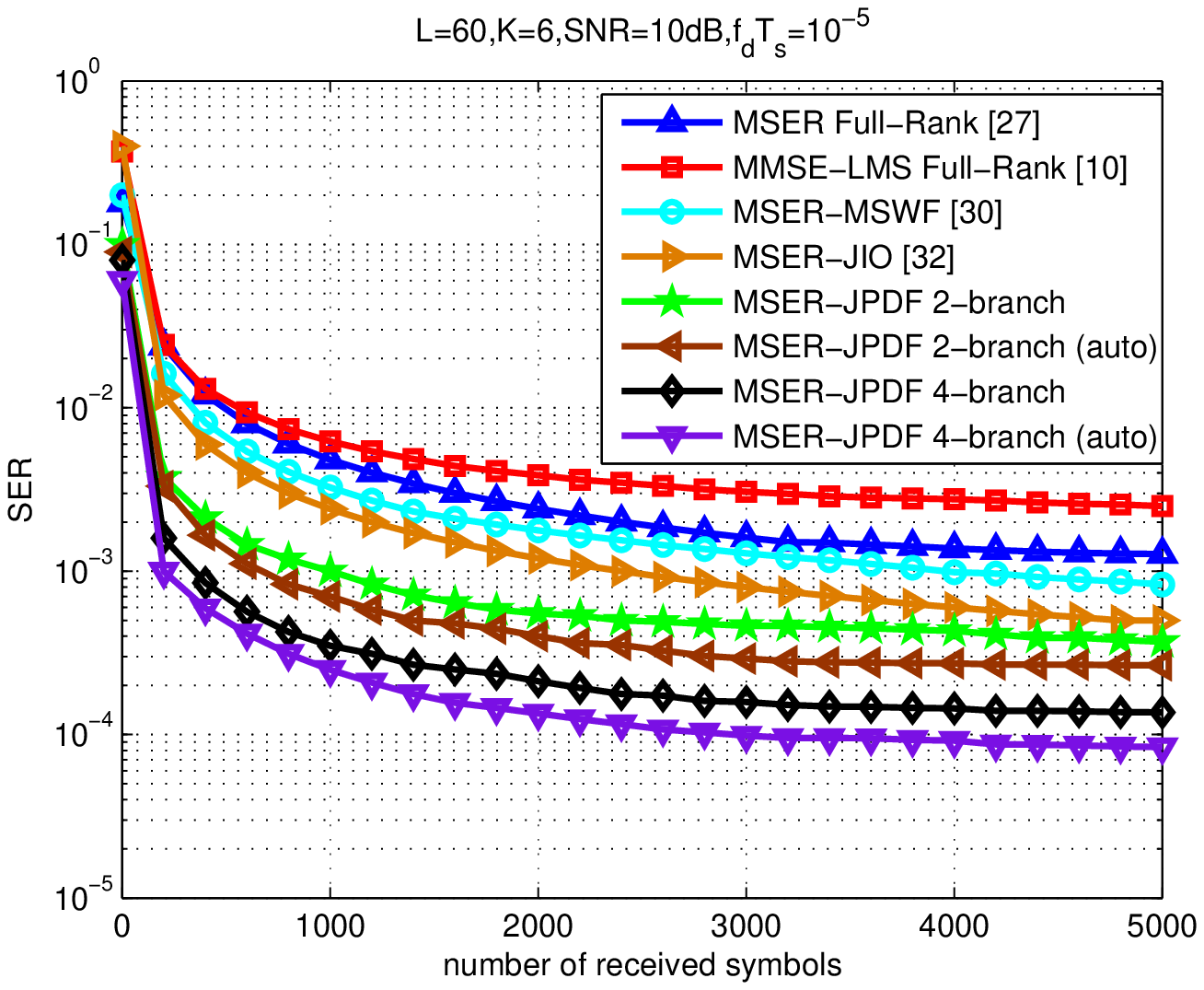}}
\caption{  SER performance of the adaptive MSER-JPDF reduced-rank algorithm with BPSK symbols
versus the number of received symbols ($K=6$,
  SNR $=10$dB,  $f_dT_s=10^{-5}$).}
\label{fig:simulation5}
\end{figure}


\begin{figure}[!hhh]
\centering
\scalebox{0.5}{\includegraphics{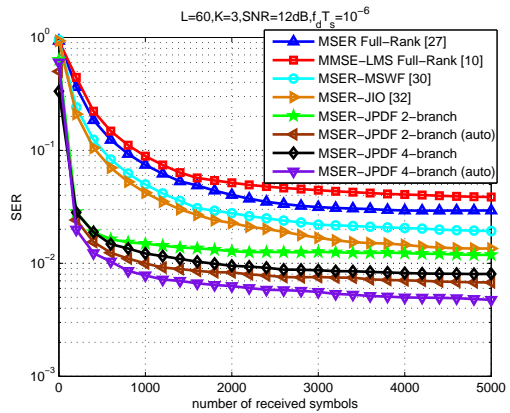}}
\caption{  SER performance of the adaptive MSER-JPDF reduced-rank algorithm with $16$-QAM symbols
versus the number of received symbols ($K=3$,
  SNR $=12$dB,  $f_dT_s=10^{-6}$).}
\label{fig:simulation6}
\end{figure}

Fig. \ref{fig:simulation5} and  Fig. \ref{fig:simulation6} show  the SER  performance versus the number of received symbols for the proposed  adaptive MSER-JPDF reduced-rank algorithms  with fixed and automatic parameter selection schemes  and for the existing adaptive full-rank and reduced-rank algorithms, namely: the MSER-SG full-rank adaptive algorithm \cite{shengchen}, \cite{shengchensp2008}, the MMSE-based LMS full-rank adaptive algorithm \cite{haykin},
the MSER-based MSWF adaptive reduced-rank algorithm \cite{mber3} and the MSER-based adaptive JIO reduced-rank algorithm \cite{caicl2013}.
%
In Fig. \ref{fig:simulation5}, we focus on the case of BPSK symbols  where the algorithm summarized in Table \ref{tab:table1} is applied.
The other system parameters are set to: $K=6$, SNR $=10$dB and $f_dT_s=10^{-5}$.
Fig. \ref{fig:simulation6} focuses on the $16$-QAM symbols where the algorithm in Table \ref{tab:table2} is applied; the other system parameters are: $K=3$, SNR $=12$dB and $f_dT_s=10^{-6}$.
The main algorithm parameters in the experiments are adjusted as given in Table \ref{tab:table5} for the BPSK case and in Table \ref{tab:table6} for the $16$-QAM case.
We note that these parameter values for the various algorithms under comparison have been optimized based on simulations.
In order to focus on the main contribution, we only consider the uncoded system in this work.
From the results, we can see that the proposed  adaptive reduced-rank  algorithms with the automatic parameter selection scheme  achieve the best convergence performance, followed by the proposed algorithms with fixed values of $D$ and $I$, the recently proposed MSER-JIO adaptive reduced-rank algorithm and the other analyzed algorithms.
 In our studies, the steady-state SER performance of the proposed reduced-rank algorithms is consistently lower than that of the conventional algorithms.
Due to the novel MSER-JPDF scheme, the proposed reduced-rank algorithms outperform the conventional algorithms in terms of both convergence behavior and steady-state performance.

\begin{table*}[t]
\centering%
\caption{\normalsize  Optimized algorithm  parameters for BPSK symbols} {
\begin{tabular}{ccc}
\hline
MMSE-LMS Full-Rank & $\mu_w=0.015$ \\
\hline
MSER Full-Rank   &  $\mu_w=0.02$ \\
\hline
 MSER-MSWF   &  $\mu_w=0.05$, $D=10$ \\
\hline
  MSER-JIO  &  $\mu_w=0.03$, $\mu_S=0.03$, $D=10$ \\
\hline
 MSER-JPDF   &  $\mu_w=0.01$, $\mu_p=0.01$, $D=10$, $I=12$  \\

 \hline
 MSER-JPDF (auto) & $\mu_w=\mu_p=0.006$, $D_{max}=10$, $D_{min}=6$, $I_{max}=12$, $I_{min}=6$\\
\hline
\end{tabular}
}\label{tab:table5}
\end{table*}

\begin{table*}[t]
\centering%
\caption{\normalsize  Optimized algorithm  parameters for $16$-QAM symbols} {
\begin{tabular}{ccc}
\hline
MMSE-LMS Full-Rank & $\mu_w=0.0015$ \\
\hline
MSER Full-Rank   &  $\mu_w=0.004$ \\
\hline
 MSER-MSWF   &  $\mu_w=0.03$, $D=12$ \\
\hline
  MSER-JIO  &  $\mu_w=0.008$, $\mu_S=0.008$, $D=12$ \\
\hline
 MSER-JPDF   &  $\mu_w=0.006$, $\mu_p=0.006$, $D=12$, $I=12$  \\

  \hline
 MSER-JPDF (auto) & $\mu_w=\mu_p=0.003$, $D_{max}=12$, $D_{min}=6$, $I_{max}=12$, $I_{min}=6$\\
\hline
\end{tabular}
}\label{tab:table6}
\end{table*}


\begin{figure}[!hhh]
\centering
\scalebox{0.42}{\includegraphics{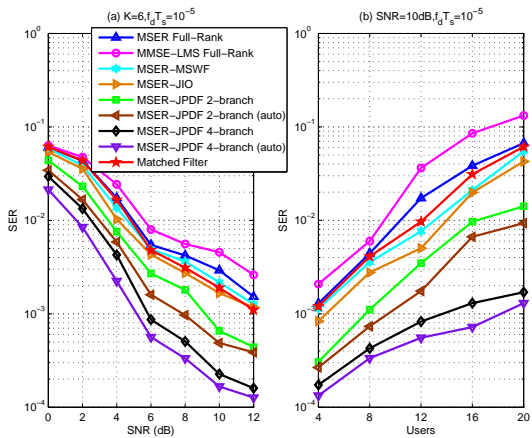}}
\caption{ SER performance for BPSK symbols
versus: (a) SNR  ($K=6$); (b) number of users $K$ (SNR $=10$ dB).}
\label{fig:simulation7}
\end{figure}


\begin{figure}[!hhh]
\centering
\scalebox{0.41}{\includegraphics{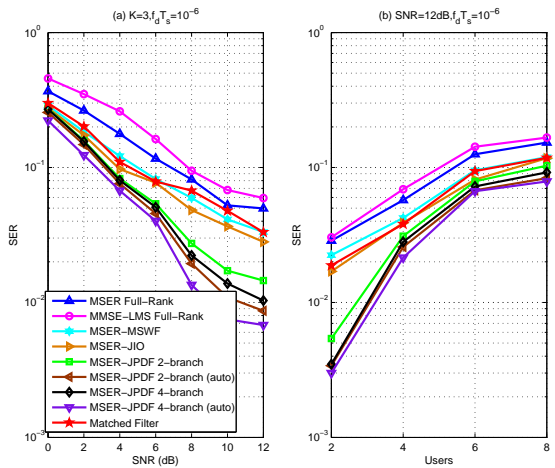}}
\caption{ SER performance for $16$-QAM symbols
versus: (a) SNR  ($K=3$); (b) number of users $K$ (SNR $=12$ dB).}
\label{fig:simulation8}
\end{figure}

 Fig. \ref{fig:simulation7}(a) and (b) show the  SER performance of the desired user versus the SNR and the number of users $K$ for BPSK symbols, respectively, where  we  set $f_dT_s=10^{-5}$. The SER is evaluated for data records of $1500$ symbols.
Fig. \ref{fig:simulation8}(a) and (b) show the corresponding results for $16$-QAM symbols, where we set $f_dT_s=10^{-6}$.
The optimized parameter values of the algorithms are shown in Table \ref{tab:table5} and Table \ref{tab:table6}.
From the results, we can see that the best performance
is achieved with the proposed adaptive MSER-JPDF reduced-rank algorithms, followed
by the adaptive MSER-JIO reduced-rank algorithm, the adaptive MSER-MSWF reduced-rank algorithm, the matched filter, the MSER full-rank adaptive algorithm and the MMSE-based LMS full-rank adaptive algorithm.
  In particular, for the BPSK case
the adaptive MSER-JPDF reduced-rank algorithm with the automatic parameter selection scheme ($2$ branches) can lead to a $4$dB gain in SNR
 and support  $5$ more users in comparison with
the adaptive MSER-MSWF reduced-rank algorithm  at the
SER level of $10^{-3}$. For the $16$-QAM case, the proposed algorithm with the automatic parameter selection scheme ($2$ branches)
  can lead to a gain more than $4$dB  in SNR and support $2$ more users compared to the adaptive MSER-JIO reduced-rank algorithm  at the
SER level of $2\times10^{-2}$.

Furthermore, we consider a  measure which relates SER and computational complexity at the same time  to compare the proposed reduced-rank algorithms with the conventional  reduced-rank algorithms.
In this scenario, we compute the packet success probability to computational complexity ratio (PCR), which is expressed as $\frac{(1-SER)^{n}}{m}$,
where $n$ denotes the number of symbols per packet and $m$ denotes the number of computations per packet. For simplicity, we focus on using the number of multiplications  as a measure of the computational complexity. Tables \ref{tab:table7} and \ref{tab:table8} show the PCR values versus different values of SNR for BPSK and $16$-QAM symbols, respectively, where we set $n=20$. The system parameters are to those used in  Fig. \ref{fig:simulation7}(a) and Fig. \ref{fig:simulation8}(a). From the results, we can see that the proposed reduced-rank algorithms provide  larger PCR values compared to the conventional reduced-rank algorithms, which indicates that  the proposed algorithms can improve the performance with reduced computational complexity.

\begin{table*}[t]
\centering%
\caption{ \normalsize  Comparison of PCR values for BPSK symbols} {
\begin{tabular}{|c|c|c|c|c|}
\hline
Algorithms & SNR $=0$dB & SNR $=4$dB & SNR $=8$dB & SNR $=12$dB \\
\hline
MSER-JPDF ($B=2$)   &  $4.52\times 10^{-6}$ & $9.54\times 10^{-6}$ & $1.07\times 10^{-5}$ & $1.1\times 10^{-5}$ \\
\hline
MSER-JPDF ($B=4$) & $3.05\times 10^{-6}$ & $5.1\times 10^{-6}$ & $5.5\times 10^{-6}$ & $5.53\times 10^{-6}$ \\
\hline
 MSER-MSWF   &  $4.49\times 10^{-8}$ & $1.15\times 10^{-7}$ & $1.4\times 10^{-7}$ & $1.47\times 10^{-7}$ \\
\hline
  MSER-JIO  &  $1.12\times 10^{-6}$ & $2.74\times 10^{-6}$ & $3.19\times 10^{-6}$ & $3.292\times 10^{-6}$ \\
\hline
%
\end{tabular}
}\label{tab:table7}
\end{table*}

\begin{table*}[t]
\centering%
\caption{  \normalsize  Comparison of PCR values for $16$-QAM symbols} {
\begin{tabular}{|c|c|c|c|c|}
\hline
Algorithms & SNR $=0$dB & SNR $=4$dB & SNR $=8$dB & SNR $=12$dB \\
\hline
MSER-JPDF ($B=2$)   &  $9.0\times 10^{-9}$ & $9.21\times 10^{-7}$ & $2.98\times 10^{-6}$ & $3.87\times 10^{-6}$\\
\hline
MSER-JPDF ($B=4$) & $5.0\times 10^{-9}$ & $4.86\times 10^{-7}$ & $1.65\times 10^{-6}$ & $2.11\times 10^{-6}$ \\
\hline
 MSER-MSWF   &  $1.4\times 10^{-10}$ & $9.37\times 10^{-9}$ & $3.67\times 10^{-8}$ & $6.35\times 10^{-8}$ \\
\hline
  MSER-JIO  &  $3.0\times 10^{-9}$ & $2.88\times 10^{-7}$ & $8.27\times 10^{-7}$ & $1.26\times 10^{-6}$ \\
\hline
%
\end{tabular}
}\label{tab:table8}
\end{table*}

\begin{figure}[!hhh]
\centering
\scalebox{0.54}{\includegraphics{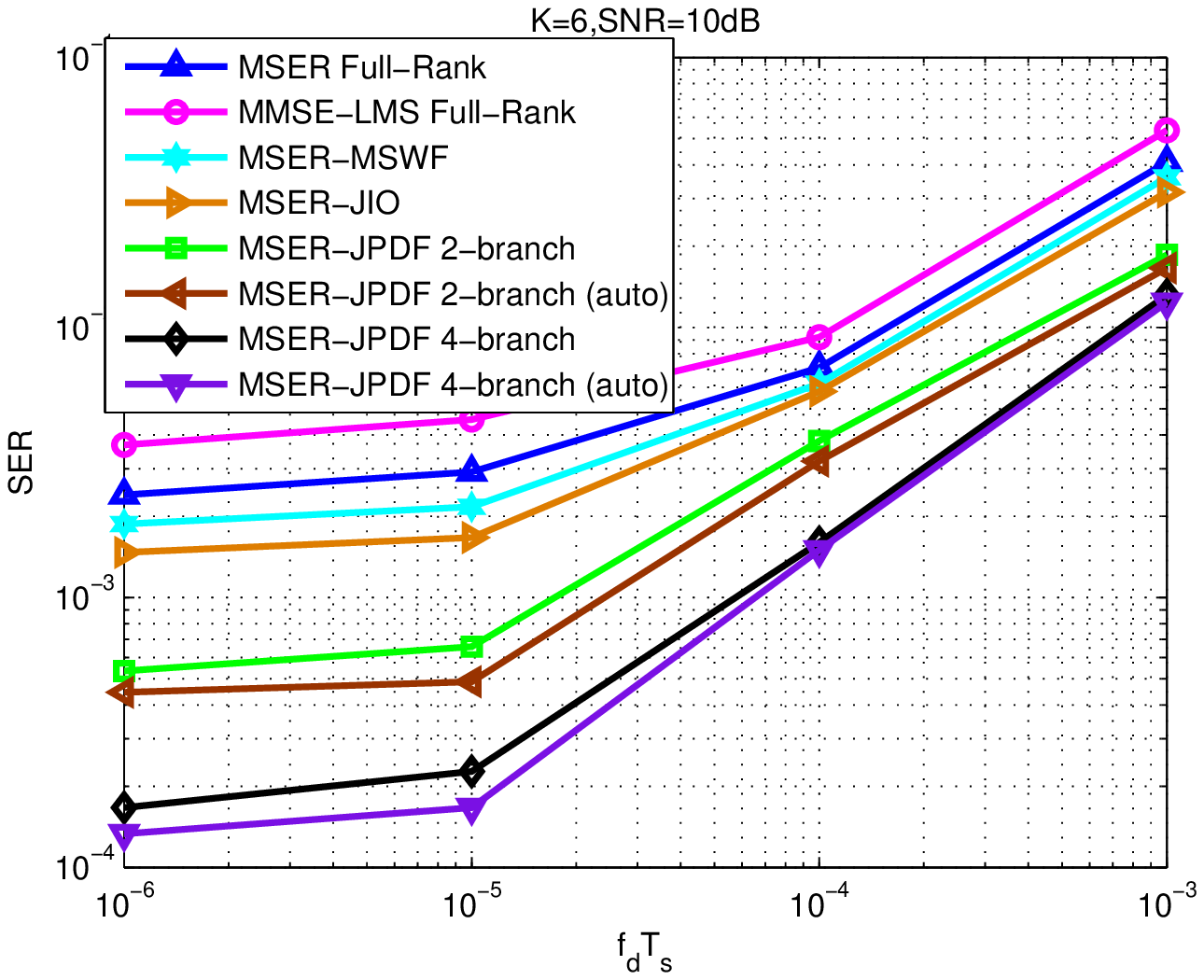}}
\caption{ SER performance
versus $f_dT_s$ (cycles/symbol) for BPSK symbols
 (SNR $=10$dB, $K=6$).}
\label{fig:simulation9}
\end{figure}

\begin{figure}[!hhh]
\centering
\scalebox{0.54}{\includegraphics{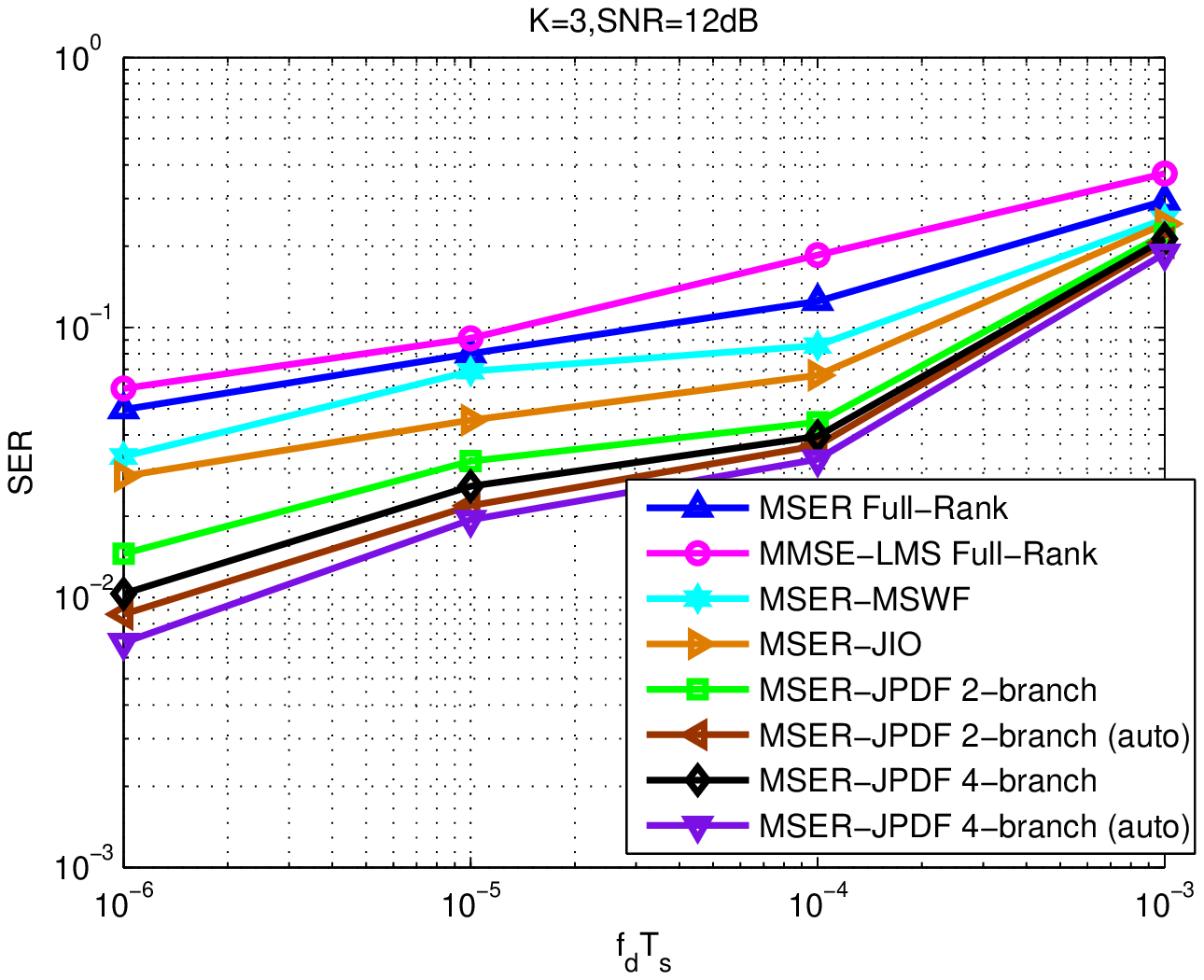}}
\caption{ SER performance
versus $f_dT_s$ (cycles/symbol) for $16$-QAM symbols
 (SNR $=12$dB, $K=3$).}
\label{fig:simulation10}
\end{figure}

In Fig. \ref{fig:simulation9} and Fig. \ref{fig:simulation10}, we show the  SER performance of the analyzed adaptive algorithms as the fading rate of the channel varies, where we employ data records of $1500$ symbols.
  The results of Fig. \ref{fig:simulation9} are based on BPSK symbols,  where we set  $K=6$ and SNR $=10$dB, while
  Fig. \ref{fig:simulation10} focuses on the $16$-QAM case, where $K=3$ and SNR $=12$dB.
The values of algorithm step-sizes used in these experiments are optimized for each value of $f_dT_s$.
 In particular, for the proposed algorithms with BPSK, we use $\mu_w=\mu_p=0.01, 0.01, 0.005, 0.002$ for $f_dT_s=10^{-6}, 10^{-5}, 10^{-4}, 10^{-3}$, respectively;
 for the $16$-QAM case, we use $\mu_w=\mu_p=0.006, 0.001, 0.0005, 0.0005$ for $f_dT_s=10^{-6}, 10^{-5}, 10^{-4}, 10^{-3}$, respectively.
First, we can see that as the fading rate increases, the
performance becomes worse, although our proposed algorithms  outperform
the existing algorithms. Moreover, we observe that the
proposed algorithms with $4$ branches perform better than in the case with $2$ branches.
Fig. \ref{fig:simulation9} and Fig. \ref{fig:simulation10}
show the ability of the adaptive MSER-JPDF reduced-rank algorithms
 to deal with channel variations for both BPSK and QAM symbols.
  Despite their low complexity, the full-rank SG-based algorithms can only achieve a poor performance, inadequate for practical applications.
 The SER performance of the analyzed adaptive algorithms in Fig. \ref{fig:simulation10} is not as good as  in Fig. \ref{fig:simulation9}, since detecting a high order modulation symbol is harder than detecting a low order modulation symbol, specially for  large values of  $f_dT_s$.



%

%
\section{Conclusion}
\label{Section7:conclusion}

In this paper, we have proposed an adaptive JPDF reduced-rank  strategy based on  the MSER criterion for the design of a receive-processing front-end in  multiuser large-scale multiple-antenna systems.
The proposed scheme employs a multiple-branch processing structure which
adaptively performs dimensionality reduction using a set of jointly optimized
preprocessing and decimation units, followed by receive filtering. The final
decision is switched to the branch with the best performance based on the
minimum Euclidean distance during a training period. We have developed SG based
algorithms for the adaptive implementation of the preprocessing filter and the
reduced-rank receive filter in the case of BPSK and QAM symbols. An automatic
parameter selection scheme has been proposed to determine the lengths of the
preprocessor and the reduced-rank receive filter during their operation. We
focused on the computational complexity and convergence to perform the analysis
of the proposed algorithms. The results have shown that the proposed scheme
significantly outperforms existing reduced-rank algorithms and can support
communication systems with higher loads, i.e. larger number of mobile users for
a given quality of services. Future work will consider nonlinear detectors and
other multiple-antenna configurations \cite{vblast}-\cite{zhou}.



\end{document}